\documentclass[aps,prd,twocolumn,amsmath,amssymb,amsfonts,nofootinbib,altaffilletter]{revtex4-1}
\usepackage{graphicx}
\usepackage{longtable}
\usepackage{hyperref}
\hypersetup{
  bookmarksopen=true
}
\usepackage{amssymb}
\usepackage{amsmath}
\usepackage{color}
\usepackage{supertabular}
\usepackage{dcolumn}
\usepackage{ wasysym }

\def\EatH{Einstein@Home}

\def\sci#1#2{#1\times10^{#2}}

\newcommand{\avgSeg}[1]{\overline{#1}}			

\newcommand{\cF}{c_*}

\newcommand{\Gauss}{\mathrm{\MakeUppercase{G}}}
\newcommand{\Signal}{{\mathrm{\MakeUppercase{S}}}}
\newcommand{\Line}{{\mathrm{\MakeUppercase{L}}}}
\newcommand{\Noise}{{\Gauss\Line}}

\providecommand{\sc}[1]{\widehat{#1}}
\renewcommand{\sc}[1]{\widehat{#1}}

\newcommand{\OSNsc}{\sc{O}_{{\Signal\Noise}}}	

\newcommand{\F}{\mathcal{F}}		

\newcommand{\avF}{\avgSeg{\F}}

\newcommand{\Nseg}{{N_{\mathrm{seg}}}}

\newcommand{\maxfdot}{2.6 $\times 10^{-9}$ Hz/s}

\newcommand{\NcandsFromStageZero}{$3.8\times 10^{10}$ }

\newcommand{\NcandsFromStageZeroAboveThr}{21.6~million }

\newcommand{\NFUOne}{16.23 $\times 10^6$ }
\newcommand{\NFUOneApprox}{16 million }
\newcommand{\NFUTwo}{5.3 $\times 10^6$ }
\newcommand{\NFUTwoApprox}{5 million }
\newcommand{\NFUThree}{1.1 million }
\newcommand{\NFUThreeApprox}{1 million }
\newcommand{\NFUFour}{10 }

\newcommand{\TcohFUZero}{60 }
\newcommand{\TcohFUOne}{60 }
\newcommand{\TcohFUTwo}{140 }
\newcommand{\TcohFUThree}{140 }
\newcommand{\TcohFUFour}{280 }

\newcommand{\NsegFUZero}{90 }
\newcommand{\NsegFUOne}{90 }
\newcommand{\NsegFUTwo}{44 }
\newcommand{\NsegFUThree}{44 }
\newcommand{\NsegFUFour}{22 }

\newcommand{\dfFUZero}{1.6\times 10^{-6} }
\newcommand{\dfFUOne}{3.6  \times 10^{-6} }
\newcommand{\dfFUTwo}{2.0  \times 10^{-6} }
\newcommand{\dfFUThree}{1.8  \times 10^{-6} }
\newcommand{\dfFUFour}{ 1.9  \times 10^{-7} }

\newcommand{\dfdotFUZero}{5.8 \times 10^{-11}}
\newcommand{\dfdotFUOne}{1\times 10^{-10}}
\newcommand{\dfdotFUTwo}{2.4 \times 10^{-11}}
\newcommand{\dfdotFUThree}{2.1 \times 10^{-11}}
\newcommand{\dfdotFUFour}{7.0 \times 10^{-12}}

\newcommand{\mskyFUZero}{0.3 + equatorial patch}
\newcommand{\mskyFUOne}{0.0042}
\newcommand{\mskyFUTwo}{0.0004}
\newcommand{\mskyFUThree}{1\times 10^{-5}}
\newcommand{\mskyFUFour}{4\times 10^{-7}}

\newcommand{\DeltafFUOne}{\pm 6.7\times 10^{-4}}
\newcommand{\DeltafFUTwo}{\pm 1.9\times 10^{-4}}
\newcommand{\DeltafFUThree}{\pm 5\times 10^{-5}}
\newcommand{\DeltafFUFour}{\pm 4 \times 10^{-7}}

\newcommand{\DeltafdotFUTwo}{\pm 3.5\times 10^{-11}}
\newcommand{\DeltafdotFUOne}{\pm 1.8\times 10^{-10}}
\newcommand{\DeltafdotFUThree}{\pm 7\times 10^{-12}}
\newcommand{\DeltafdotFUFour}{\pm 4.0\times 10^{-13}}

\newcommand{\DeltaskyFUOne}{0.55}
\newcommand{\DeltaskyFUTwo}{0.19}
\newcommand{\DeltaskyFUThree}{0.4}
\newcommand{\DeltaskyFUFour}{0.03}

\newcommand{\TotCFUOne}{4 }

\newcommand{\TotWUsFUOne}{\NFUOne}
\newcommand{\TotWUsFUTwo}{\NFUTwo}
\newcommand{\TotWUsFUThree}{\NFUThree }

\newcommand{\WUCFUOne}{2 }
\newcommand{\WUFUTwo}{12 }
\newcommand{\WUCFUThree}{2 }
\newcommand{\WUCFUFour}{14 }

\newcommand{\ThrFUZero}{6.109 }
\newcommand{\ThrFUOne}{6.109 }
\newcommand{\ThrFUTwo}{7.38 }
\newcommand{\ThrFUThree}{8.82 }
\newcommand{\ThrFUFour}{15.0 }
\newcommand{\ThrLowerCluster}{5.9} 
 
\newcommand{\fdOccupancyVeto}{$\sim$ 0.02\% of signal clusters }

\newcommand{\fdFUOne}{ $\sim$ 9\% } 
\newcommand{\fdFUTwo}{ $\sim$ 0.6\% } 
\newcommand{\fdFUThree}{$\sim 4\times 10^{-4}$  } 
\newcommand{\maxTwoFInjectionFUFour}{16.2} 

\newcommand{\nrOccupancyVeto}{45\% of signal clusters }

\newcommand{\nrFUOne}{70\%  }
\newcommand{\nrFUTwo}{79\% }
\newcommand{\nrFUThree}{99.9991\%  }
\newcommand{\nrFUFour}{$2\times 10^{-20}$} 

\newcommand{\cstarFUFour}{96.1} 

\newcommand{\ULtrialBands}{100}

\newcommand{\AverageSD}{46.9}

\newcommand{\MaxTwoFClusters}{8.6 }
\newcommand{\FreqMaxTwoFClusters}{52 Hz}
\newcommand{\SecondMaxTwoFClusters}{7.1 }

\newcommand{\MinUL}{4.3 $\times 10^{-25}$}
\newcommand{\MinFreqMin}{170.5 }
\newcommand{\MinFreqMax}{171 }

\newcommand{\ULHF}{7.6 $\times 10^{-25}$}
\newcommand{\ULuncertainty}{20\%} 

\newcommand{\epsOneHundredPcAtThreeHundredHz}{6 $\times 10^{-7}$}

\newcommand{\TrefFUfour}{960541454.5}

\newcommand{\If}{50.19985463}
\newcommand{\IIf}{50.20001612}
\newcommand{\IIIf}{52.80832455}
\newcommand{\IVf}{52.80832422}
\newcommand{\Vf}{124.60002077}
\newcommand{\VIf}{265.57623841}
\newcommand{\VIIf}{367.83543941}
\newcommand{\VIIIf}{430.28626637}
\newcommand{\IXf}{500.36312713}
\newcommand{\Xf}{500.36594568}
 
\newcommand{\Ialpha}{4.7716026}
\newcommand{\IIalpha}{4.7124554}
\newcommand{\IIIalpha}{5.2805366}
\newcommand{\IValpha}{5.2819543}
\newcommand{\Valpha}{4.7067880}
\newcommand{\VIalpha}{1.2487972}
\newcommand{\VIIalpha}{1.4807437}
\newcommand{\VIIIalpha}{6.1499768}
\newcommand{\IXalpha}{4.7121294}
\newcommand{\Xalpha}{4.5662765}

\newcommand{\Idelta}{1.1412922}
\newcommand{\IIdelta}{1.1683832}
\newcommand{\IIIdelta}{-1.4631895}
\newcommand{\IVdelta}{-1.4632398 }
\newcommand{\Vdelta}{1.1648704}
\newcommand{\VIdelta}{-0.9812202}
\newcommand{\VIIdelta}{0.7112582}
\newcommand{\VIIIdelta}{0.9203753}
\newcommand{\IXdelta}{1.1617860}
\newcommand{\Xdelta}{1.4276343}

\newcommand{\Ifdot}{3.013 $\times 10^{-11}$}
\newcommand{\IIfdot}{-5.674 $\times 10^{-12}$}
\newcommand{\IIIfdot}{7.311 $\times 10^{-14}$}
\newcommand{\IVfdot}{2.968 $\times 10^{-14}$}
\newcommand{\Vfdot}{-4.164 $\times 10^{-12}$}
\newcommand{\VIfdot}{-4.015 $\times 10^{-12}$}
\newcommand{\VIIfdot}{-9.236 $\times 10^{-10}$}
\newcommand{\VIIIfdot}{-2.056 $\times 10^{-9}$}
\newcommand{\IXfdot}{9.878 $\times 10^{-13}$}
\newcommand{\Xfdot}{-2.507 $\times 10^{-9}$}

\newcommand{\ITwoF}{11.6}
\newcommand{\IITwoF}{12.3}
\newcommand{\IIITwoF}{52.0}
\newcommand{\IVTwoF}{55.9}
\newcommand{\VTwoF}{11.8}
\newcommand{\VITwoF}{37.3}
\newcommand{\VIITwoF}{10.4}
\newcommand{\VIIITwoF}{10.0}
\newcommand{\IXTwoF}{12.2}
\newcommand{\XTwoF}{10.6}

\newcommand{\ITwoFH}{ 6.9 }
\newcommand{\IITwoFH}{ 5.5 }
\newcommand{\IIITwoFH}{ 16.9 }
\newcommand{\IVTwoFH}{ 18.1 }
\newcommand{\VTwoFH}{ 11.2}
\newcommand{\VITwoFH}{ 25.1 }
\newcommand{\VIITwoFH}{ 9.5}
\newcommand{\VIIITwoFH}{ 7.3}
\newcommand{\IXTwoFH}{ 11.9 }
\newcommand{\XTwoFH}{ 10.0 }

\newcommand{\ITwoFL}{9.5}
\newcommand{\IITwoFL}{11.2 }
\newcommand{\IIITwoFL}{39.7 }
\newcommand{\IVTwoFL}{44.0 }
\newcommand{\VTwoFL}{ 6.1}
\newcommand{\VITwoFL}{17.0}
\newcommand{\VIITwoFL}{4.9}
\newcommand{\VIIITwoFL}{5.5}
\newcommand{\IXTwoFL}{5.4}
\newcommand{\XTwoFL}{4.6}

\newcommand{\IIIinjf}{52.8083244}
\newcommand{\VIinjf}{265.5762386}

\newcommand{\IIIinjalpha}{5.281831296}
\newcommand{\VIinjalpha}{1.248816734}

\newcommand{\IIIinjdelta}{-1.463269033}
\newcommand{\VIinjdelta}{-0.981180225}

\newcommand{\IIIinjfdot}{-4.03 $\times 10^{-18}$}
\newcommand{\VIinjfdot}{-4.15 $\times 10^{-12}$}

\newcommand{\IIIDeltaf}{1.5 $\times 10^{-7}$}
\newcommand{\IIIDeltaalpha}{-1.29 $\times 10^{-3}$}
\newcommand{\IIIDeltadelta}{7.95 $\times 10^{-5}$}
\newcommand{\IIIDeltafdot}{7.3 $\times 10^{-14}$}

\newcommand{\IVDeltaf}{-1.8 $\times 10^{-7}$}
\newcommand{\IVDeltaalpha}{1.23 $\times 10^{-4}$}
\newcommand{\IVDeltadelta}{2.92 $\times 10^{-5}$}
\newcommand{\IVDeltafdot}{3.0 $\times 10^{-14}$}

\newcommand{\VIDeltaf}{-1.9 $\times 10^{-7}$}
\newcommand{\VIDeltaalpha}{-1.95 $\times 10^{-5}$}
\newcommand{\VIDeltadelta}{-4.00 $\times 10^{-5}$}
\newcommand{\VIDeltafdot}{1.4 $\times 10^{-13}$}

\let\svthefootnote\thefootnote

\begin{document}

\title{ 
Hierarchical follow-up of sub-threshold candidates of an all-sky Einstein@Home search for continuous gravitational waves on LIGO sixth science run data
}
\author{Maria Alessandra Papa$^\mathrm{1,2,4,a}$, Heinz--Bernd Eggenstein$^\mathrm{2,3}$, Sin\'ead Walsh$^\mathrm{1,2}$, Irene Di Palma$^\mathrm{1,2,5}$, Bruce Allen$^\mathrm{2,4,3}$, Pia Astone$^\mathrm{5}$, Oliver Bock$^\mathrm{2,3}$, Teviet D. Creighton$^\mathrm{7}$, David Keitel$^\mathrm{2,3,6}$, Bernd Machenschalk$^\mathrm{2,3}$, Reinhard Prix$^\mathrm{2,3}$, Xavier Siemens$^\mathrm{4}$, Avneet Singh$^\mathrm{1,2,3}$, Sylvia J. Zhu$^\mathrm{1,2}$, Bernard F. Schutz,$^{8,1}$\\\vspace{0.3in}
}\let\thefootnote\relax\footnote{\textsuperscript{a}email: maria.alessandra.papa@aei.mpg.de}
\affiliation{$^1$ Max-Planck-Institut f{\"u}r Gravitationsphysik, am M{\"u}hlenberg 1, 14476 Potsdam, Germany\\ 
$^2$ Max-Planck-Institut f{\"u}r Gravitationsphysik, Callinstra{$\beta$}e 38, 30167 Hannover, Germany\\ 
$^3$ Leibniz Universit{\"a}t Hannover, Welfengarten 1, 30167 Hannover, Germany \\
$^4$ University of Wisconsin-Milwaukee, Milwaukee, Wisconsin 53201, USA \\
$^5$ Universita di Roma ``La Sapienza'', P.zle A. Moro 2, 00185 Roma, Italy\\
$^6$ Universitat de les Illes Balears, IAC3---IEEC, E-07122 Palma de Mallorca, Spain\\
$^7$ The University of Texas Rio Grande Valley, Brownsville, TX 78520, USA\\
$^8$ Cardiff University, Cardiff CF24 3AA, United Kingdom\\\vspace{0.1in}}
\addtocounter{footnote}{-1}\let\thefootnote\svthefootnote


\begin{abstract}
We report results of an all-sky search for periodic gravitational waves with frequency between 50 and 510 Hz from isolated compact objects, e.g. neutron stars. A new hierarchical multi-stage approach is taken, supported by the computing power of the \EatH~ project, allowing  to probe more deeply than ever before. 16 million sub-threshold candidates from the initial search \cite{S6BucketStage0} are followed up in three stages. None of those candidates is consistent with an isolated gravitational wave emitter,
and 90\%\ confidence level upper limits are placed on the amplitudes of continuous waves from the target population. 
Between \MinFreqMin and \MinFreqMax ~Hz we set the most constraining 90\%\ confidence upper limit on the strain amplitude $h_0$ at \MinUL, while at the high end of our frequency range we achieve an upper limit of \ULHF. These are the most constraining all-sky upper limits to date and constrain the ellipticity of rotating compact objects emitting at $300$ Hz at a distance $D$ to less than \epsOneHundredPcAtThreeHundredHz $\left[ D\over {100 ~ \textrm{pc}} \right]$.
\end{abstract}

\preprint{LIGO-P}
\maketitle

\section{Introduction}
\label{sec:introduction}


The beauty of continuous signals is that, even if a candidate is not significant enough to be recognised as a real signal after a first semi-coherent search, it is still possible to improve its significance to the level necessary to claim a detection after a series of follow-up searches. Hierarchical approaches were first proposed in the late 90s and developed over a number of searches on LIGO data: \cite{FullS5EH} and \cite{Shaltev:2014toa} detail a semi-coherent search plus a three-stage follow-up of order 100 candidates; \cite{GalacticCenterSearch} and \cite{GalacticCenterMethod} detail a semi-coherent search plus a series of vetoes and a final coherent follow-up of over 1000 candidates. The search detailed here follows up 16 million candidates and is the first large-scale hierarchical search ever done.

We use a hierarchical approach consisting of 4 stages applied to the processed results (``Stage 0'') of an initial search \cite{S6BucketStage0}.  At each stage a semi-coherent search is performed and the top ranking cells in parameter space (also referred to as ``candidates'') are marked and are searched in the next stage. At each stage the significance of a cell harbouring a real signal would increase with respect to the significance it had in the previous stage. The significance of a cell that did not contain a signal on the other hand is not expected to increase consistently over the different stages. In the first three stages the thresholds that define the top ranking cells are low enough that many false alarms are expected over the large parameter space that was searched. And indeed at the end of the first stage we have \NFUOneApprox $~$ candidates. At the end of the second stage we have \NFUTwoApprox. At the end of the third stage we have \NFUThreeApprox. At the end of the fourth stage we are left with only \NFUFour $~$ candidates. 

The paper is organised very simply: Section \ref{sec:method} introduces the quantities that characterise each stage of the follow up. Section \ref{sec:search} illustrates how the different stages were set up for the S6 LIGO \EatH~ candidates follow-ups. Section \ref{sec:results} present the gravitational wave amplitude and ellipticity upper limit results. In the last section, Section \ref{sec:conclusions}, we summarise the main findings and discuss prospects for this type of search.

\section{Quantities defining each stage}
\label{sec:method}

From one stage to the next in this hierarchical scheme, the number of surviving candidates is reduced, the uncertainty over the signal parameters for each candidate is also reduced and the significance of a real signal increased. This latter effect is due both to search being intrinsically more sensitive and to the trials' factor decreasing for every search from one stage to the next.

Each stage performs a stack-slide type of search using the GCT method and implementation of \cite{Pletsch:2008,Pletsch:2010}. Important variables are: the coherent time baseline of the segments, the number of segments used ($\Nseg$), the total time spanned by the data, the grids in parameter space and the detection statistic used to rank the parameter space cells. All stages use the same data set. The first three follow-up searches are performed on the Einstein@Home volunteer computing platform \cite{EaHweb}, the last on the Atlas computing cluster \cite{Atlas}.
\begin{figure}[h!tbp]
   \includegraphics[width=\columnwidth]{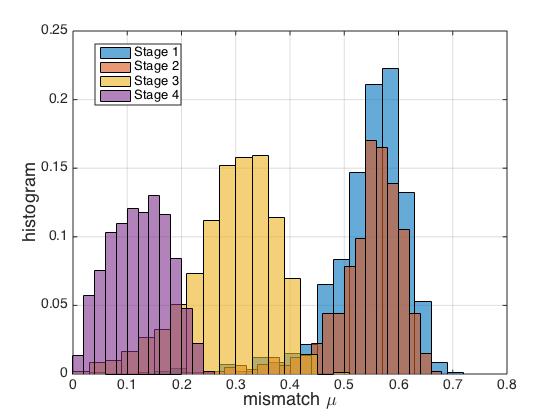}
\caption{These are the mismatch histograms of the four follow-up searches, so the y-axis represents normalised counts. For a given search and search set-up, the mismatch distribution depends of the template grid. The injection-and-recovery Monte Carlo studies to determine these distributions were performed without noise.}  
\label{fig:mismatch}
\end{figure}

The parameters for the various stages are summarised in Table \ref{tab:GridSpacings}. The grids in frequency and spindown are each described by a single parameter, the grid spacing, which is constant over the search range. The same frequency grid spacings are used for the coherent searches over the segments and for the incoherent summing. The spindown spacing for the incoherent summing step is finer than that used in for the coherent searches by a factor $\gamma$. The notation used here is consistent with that used in previous observational papers \cite{GalacticCenterSearch,S6BucketStage0,S5GC1HF} and in the GCT methods papers \cite{Pletsch:2008,Pletsch:2010}. 

The sky grids for stages 1 to 4 are approximately uniform on the celestial sphere projected on the ecliptic plane. 
The tiling is an hexagonal covering of the unit circle with hexagons' edge length $d$: 
\begin{equation}
d(m_{\text{sky}})={1\over f}
{
{\sqrt{ m_{\text{sky}} } }
\over {\pi \tau_{E}} 
} ,
\label{eq:skyGridSpacing}
\end{equation}
with $\tau_{E}\simeq0.021$ s being half of the light travel time across the Earth and $m_{\text{sky}}$ the so-called mismatch parameter. As was done in previous searches \cite{FullS5EH, S6BucketStage0} the sky-grids are constant over 10Hz bands and the spacings are the ones associated through Eq.~\ref{eq:skyGridSpacing} to the highest frequency in the range. The sky grid of stage 0 is the union of two grids: one is uniform on the celestial sphere after projection onto the equatorial plane and the tiling (in the equatorial plane) is approximately square with edge $d(0.3)$ from Eq.\ref{eq:skyGridSpacing}; the other grid is limited to the equatorial region ($0\leq \alpha\leq 2\pi$ and $-0.5\leq \delta \leq 0.5$), with constant actual $\alpha$ and $\delta$ spacings equal to $d(0.3)$ -- see Fig.1 of \cite{S6BucketStage0}. The reason for the equatorial ``patching'' with a denser sky grid is to improve the sensitivity of the search.

\begin{table*}[t]
\centering
\begin{tabular}{|c|c|c|c|c|c|c|}
\hline
 & $T_{coh}$ & $N_{seg}$ & $\delta f$ & $\delta {\dot{f_c}}$ & $\gamma$ & $m_{\text{sky}}$ \\
& hrs &  & Hz &  Hz/s & & \\
\hline
Stage 0 &\TcohFUZero & \NsegFUZero &   $\dfFUZero$ & $\dfdotFUZero$ &  230 & \mskyFUZero\\
\hline
Stage 1 &\TcohFUOne & \NsegFUOne & $\dfFUOne$   & $\dfdotFUOne$ &  230 & \mskyFUOne \\
\hline
Stage 2 &\TcohFUTwo & \NsegFUTwo &  $\dfFUTwo$ & $\dfdotFUTwo$& 100  & \mskyFUTwo\\
\hline
Stage 3 & \TcohFUThree & \NsegFUThree & $\dfFUThree$ & $\dfdotFUThree$ & 100  & $\mskyFUThree$\\
\hline
Stage 4 & \TcohFUFour &  \NsegFUFour &  $\dfFUFour$	& $\dfdotFUFour$ & 	50  & $\mskyFUFour$  \\\hline
\end{tabular}
\caption{Search parameters for each of the semi-coherent stages.}
\label{tab:GridSpacings}
\end{table*}

After each stage a threshold is set on the detection statistic to determine what candidates will be searched by the next stage. We set this detection threshold to be the highest such that the weakest signal that survived the first stage of the pipeline would, with high confidence, not be lost. 

The set-up for each stage is determined at fixed computation cost. The computational cost is mostly set by practical considerations such as the time-frame on which we'd like to have a result, the number of stages that we envision in the hierarchy and the availability of \EatH.

Since an analytical model that predicts the sensitivity of a search with the current implementation of the GCT method does not exist, we consider different search set-ups and for every set-up we perform fake-signal injection and recovery Monte Carlos. From these we determine the detection efficiency and the signal parameter uncertainty for signals at the detection threshold. We pick the search set-up based on these. Typically the search setup with the lowest parameter uncertainty volume ($R$ given below) also has the highest detection efficiency and we pick that. As a further cross-check we also determine the mismatch distributions for the detection statistic. We define the mismatch $\mu$ as
\begin{equation}
\label{eq:mismatch}
\mu = \frac{2\avF_{\mathrm{signal}} - 2\avF_{\mathrm{candidate}}}{2\avF_{\mathrm{signal}} - 4}
\end{equation}
where $\avF_{\mathrm{signal}}$ is the value of the detection statistic that we measure when we search the data with a template that is perfectly matched to the signal and $\avF_{\mathrm{candidate}}$ is the value of the detection statistic that we obtain when running a search on a set of templates none of which, in general, will perfectly coincide with the signal waveform. The mismatch is hence a measure of how fine the grid that we are using is. As expected, Fig.~\ref{fig:mismatch} shows that the grids of subsequent stages get finer and finer. 

At each stage we determine the signal parameter uncertainty for signals at least at the detection threshold, in each search dimension: the distance in parameter space around a candidate that with high confidence (at least 90\%), includes the signal parameter values. The uncertainty region around each candidate associated with stage $i$ is searched in stage $i+1$. The uncertainty volume at stage $i$ is smaller than the containment volume of stage $i-1$.  

\section{The S6 search follow-up}
\label{sec:search}

A series of all-sky \EatH~ searches looked for signals with  frequencies from 50~Hz through 510~Hz and frequency derivatives from $\sci{3.1}{-10}$~Hz/s through $-\sci{2.6}{-9}$~Hz/s. Results from these were combined and analysed as described in \cite{S6BucketStage0}: no significant candidate was found and upper limits were set on the gravitational wave signal amplitude in the target signal parameter space. The data set that we begin with, is that described in Section III.1 and III.2 of \cite{S6BucketStage0}: a ranked-list of \NcandsFromStageZero candidates each with an associated detection statistic value $2\avF$. We now take the \NFUOneApprox most promising regions in parameter space from that search and inspect them more closely. This is done in four stages which we describe in the next subsections.

We remind the reader that some of the input data to this search was treated by substituting the original frequency-domain data with fake Gaussian noise at the same level as that of the neighbouring frequencies. This is done in frequency regions affected by well-known artefacts, as described in \cite{S6BucketStage0}. Results stemming entirely from this fake data are not considered in any further stage. Moreover, after the initial \EatH~ search, the results in 50 mHz bands were visually inspected and those 50 mHz bands that present obvious noise disturbances are also removed from the analysis. A complete list of the excluded bands is given in the Appendixes of \cite{S6BucketStage0}. We will come back to this point as we present the results of this search.

\subsection{Stage 0}
\label{subsec:stage0}
This is the most complex stage of the hierarchy and determines the sensitivity of the search: if a signal does not pass this initial stage it will be lost forever.  
So we try here to keep the threshold that candidates have to exceed to be considered further as low as possible, compatibly with the feasibility of the next stage with the available computing resources. Such threshold was set at $2\avF=\ThrFUZero$. 

\begin{figure}[h!tbp]
   \includegraphics[width=\columnwidth]{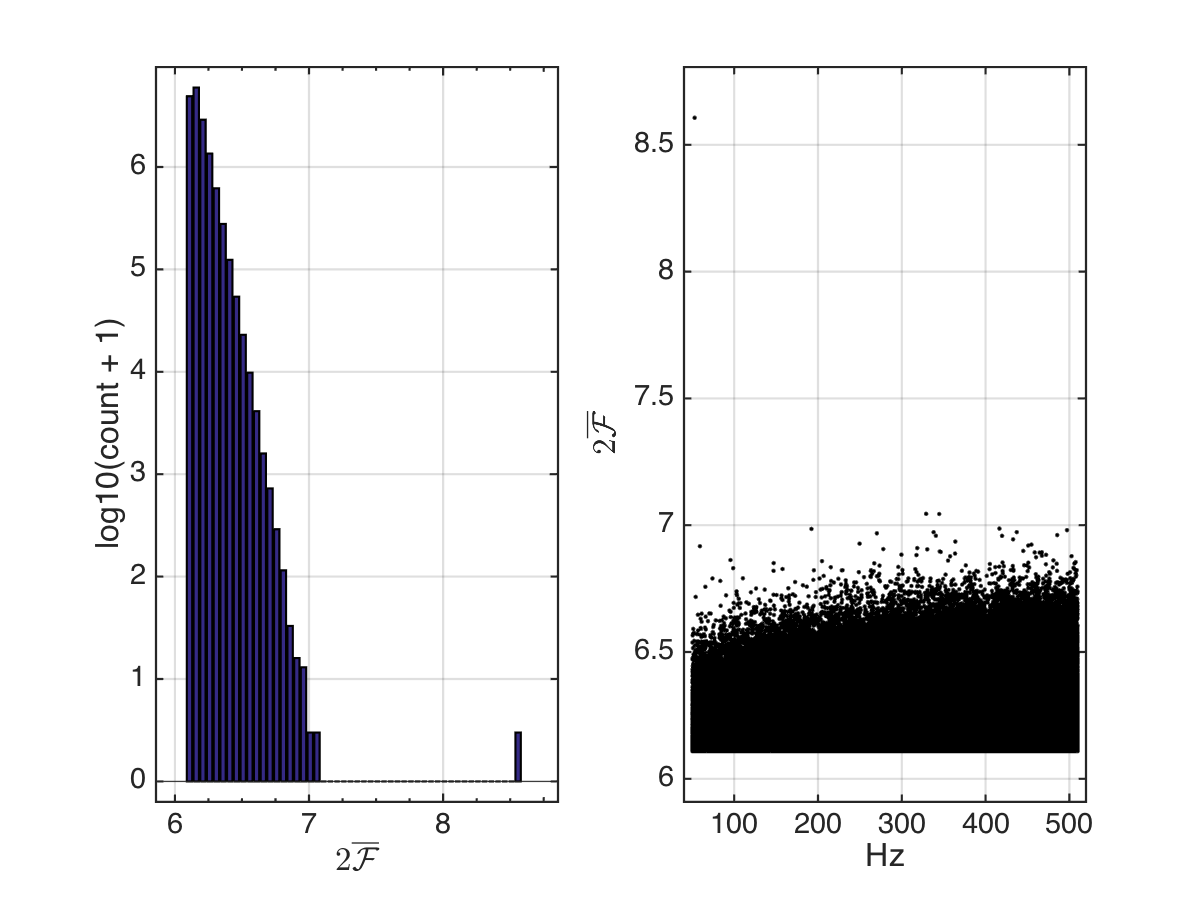}
\caption{Candidates that are followed-up in stage 1 : the distribution of their detection statistic values $2\avF$ (left plot) and their distribution as a function of frequency (right plot).}  
\label{fig:ClusterSeeds}
\end{figure}

The identification of correlated candidates saves compute cycles in the next steps of the search. As was done in \cite{GalacticCenterSearch} the clustering procedure aims at bundling together candidates that could be ascribed to the same signal. In fact, a loud signal as well as a loud disturbance would produce high values of the detection statistic at a number of different template grid points and it would be a waste to follow up each of these independently. 
As described in \cite{GalacticCenterSearch,GalacticCenterMethod} we begin with the loudest candidate, i.e. the candidate with the highest value of $2\avF$ . This is the seed for the first cluster. We associate with it close-by candidates in parameter space. Together, the seed and the nearby candidates, constitute the first cluster. We remove the candidates from the first cluster from the candidate list. The loudest candidate on the resulting list is the seed of the second cluster. We proceed in the same way as for the first cluster and reiterate the procedure until no more seeds with $2\avF$ values equal to or larger than \ThrFUZero remain.

Monte Carlo studies are conducted to determine the cluster box size, i.e. the neighbourhood of the seed that determines the cluster occupants. We inject signals in Gaussian noise data at the level of our detectors' noise, search a small parameter space region around the signal parameters and use the resulting candidates as representative of what we would find in an actual search, after comparing them with the candidates obtained in the case of 'noise-only'. For signals at the detection threshold the 90\% confidence cluster box is:
\begin{equation}
  \begin{cases}
    \Delta f ^{\text{ Stage-0}}&= \pm 1.2\times 10^{-3} {~\text{Hz}}\\
    \Delta {\dot{f}} ^{\text{ Stage-0}}&= \pm 2.6\times 10^{-10} {~\text{Hz/s}}\\
    \Delta {\text{sky}} ^{\text{ Stage-0}}&\simeq 25 {\text{ points around seed}}
      \end{cases}
      \label{eq:clusterSize}
\end{equation}

If we consider as cluster occupants only those with $2\avF$ values greater or equal to \ThrLowerCluster, we observe that signals tend to produce slight over-densities in the clusters with respect to noise. This feature is exploited with an {\it{occupancy veto}} that discards all clusters with less than 2 occupants. We find that the false dismissal for signals at threshold is hardly affected (\fdOccupancyVeto) whereas the noise rejection is quite significant: we exclude \nrOccupancyVeto.

This same set of injection-data is utilised to characterise the false dismissals and the parameter uncertainty regions for all the stages of the hierarchy.

To summarise: the total number of candidates returned by the \EatH~ searches is \NcandsFromStageZero. Of these we consider the ones with $2\avF$ above \ThrFUZero, excluding frequency bands with obvious noise disturbances. There are \NcandsFromStageZeroAboveThr such candidates. After clustering and occupancy veto we have reduced this number to \NFUOne. The distribution of the detection statistic values $2\avF$ for these candidates is shown in Fig. \ref{fig:ClusterSeeds} as well as their distribution in frequency. The maximum value is \MaxTwoFClusters and occurs at $\sim$ \FreqMaxTwoFClusters. All remaining values are smaller than \SecondMaxTwoFClusters.

\subsection{Stage 1}
\label{subsec:stage1}

In this stage we search a volume of parameter space (Eqs. \ref{eq:clusterSize}) around each candidate (around each cluster seed) equal to the cluster box defined in \ref{eq:clusterSize}. We fix the total run time to be \TotCFUOne months on \EatH~ and this yields an optimal search set-up having the same coherent time baseline as stage 0, \TcohFUOne hours, with the same number of segments $N_{seg}=\NsegFUOne$ and the grid spacings shown in Table \ref{tab:GridSpacings}. We use the same ranking statistic as in the original search \cite{S6BucketStage0}, the $\OSNsc$ \cite{Keitel:2013}, with the same tunings. The 90\% uncertainty regions for this search set-up for signals just above the detection threshold are 
\begin{equation}
  \begin{cases}
    \Delta f ^{\text{ Stage-1}}&= \DeltafFUOne {~\text{Hz}}\\
    \Delta {\dot{f}} ^{\text{ Stage-1}}&= \DeltafdotFUOne {~\text{Hz/s}}\\
    \Delta {\text{sky}} ^{\text{ Stage-1}}&\simeq  \DeltaskyFUOne ~\Delta {\text{sky}} ^{\text{ Stage-0}}
      \end{cases}
      \label{eq:FU2Box}
\end{equation}

\begin{figure}[h!tbp]
   \includegraphics[width=\columnwidth]{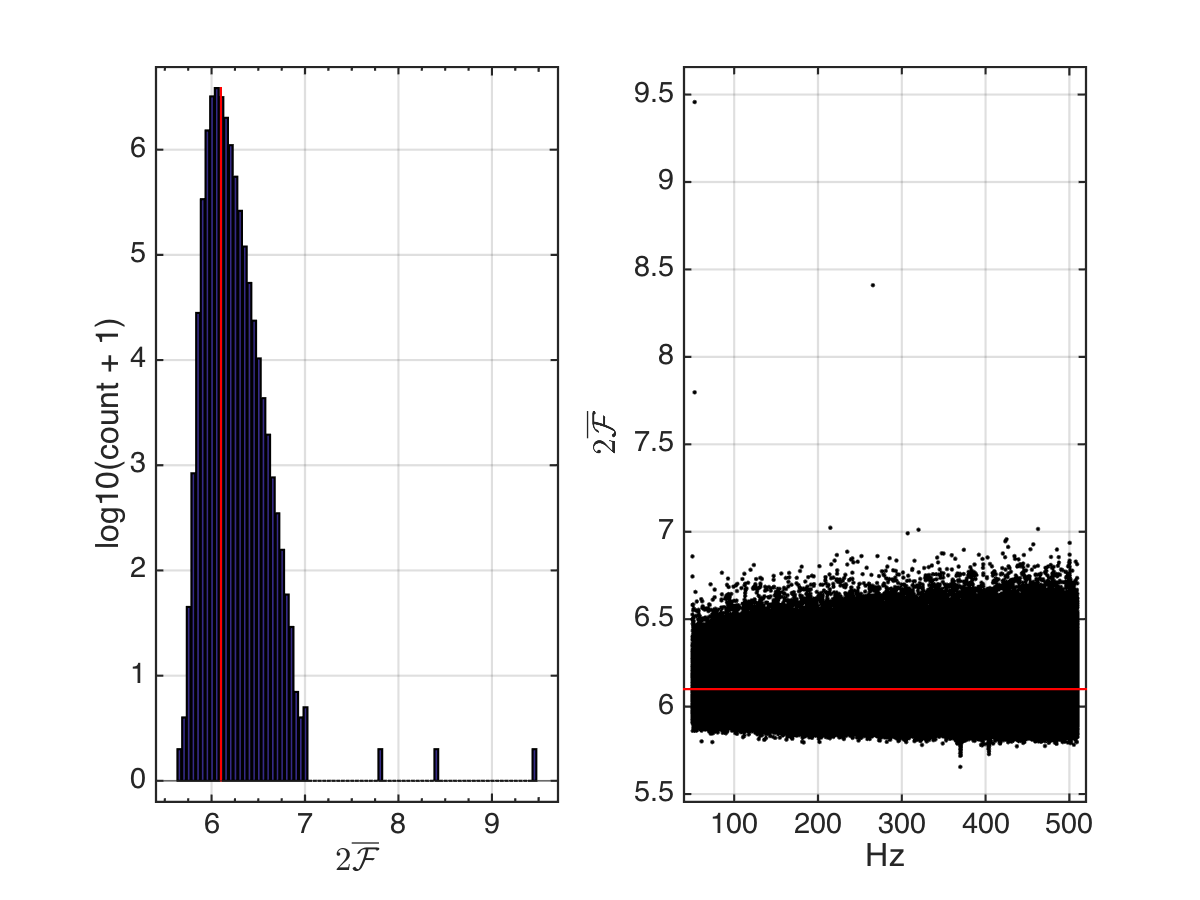}
\caption{Loudest from each of the stage-1 searches: the distribution of their detection statistic values $2\avF$ (left plot) and their distribution as a function of frequency (right plot). The red line marks $2\avF=\ThrFUOne$ which is the threshold at and above which candidates are passed on to stage-2.}  
\label{fig:FU1Out}
\end{figure}

The search is divided among \TotWUsFUOne work-units (WUs) each lasting about \WUCFUOne hours and performed by one of the \EatH volunteer computers. From each follow-up search we record the most significant candidate. The distribution of these is shown in Fig.\ref{fig:FU1Out}. A threshold at $\avF=$  \ThrFUOne has a \fdFUOne false dismissal for signals at threshold and a \nrFUOne noise rejection. Using this threshold to determine what candidates to consider in the next stage yields \NFUTwo candidates. 

\begin{figure}[h!tbp]
   \includegraphics[width=\columnwidth]{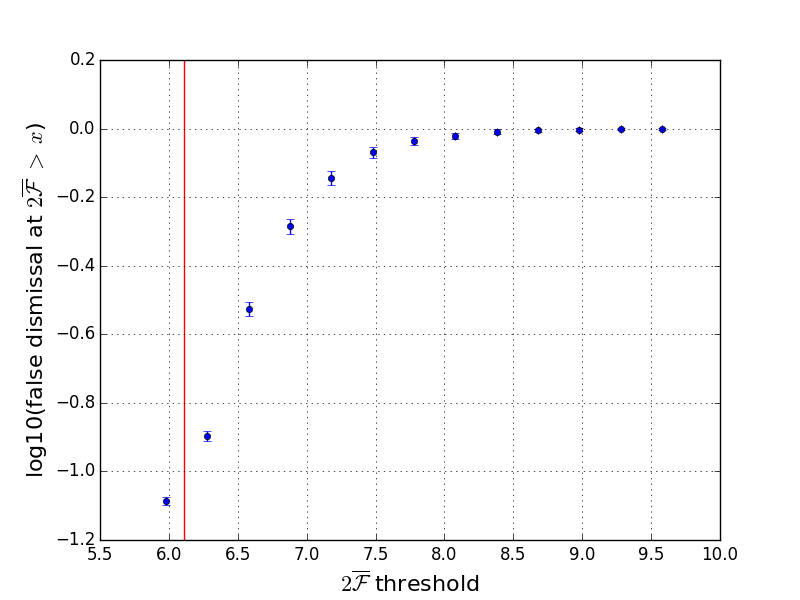}
\caption{Fraction of signals that are recovered with a detection statistic value larger than or equal to the threshold value after the stage-1 follow-up.}  
\label{fig:FU1FD}
\end{figure}

\subsection{Stage 2}
\label{subsec:stage2}

In this stage we search a volume of parameter space around each candidate defined by Eq. \ref{eq:FU2Box}. 
As shown in Table \ref{tab:GridSpacings}, we use a coherent time baseline which is about twice as long as that used in the previous stages and the grid spacings finer. The ranking statistic is $\OSNsc$ with the same tunings ($\cF$ and normalized SFT power threshold) as in the previous stages. 
The computational load is divided among \TotWUsFUTwo WUs, each lasting about \WUFUTwo hrs.

The $>$ 99\% uncertainty regions for this search set-up for signals close to the detection threshold are 
\begin{equation}
  \begin{cases}
    \Delta f ^{\text{ Stage-2}}&= \DeltafFUTwo {~\text{ Hz}}\\
    \Delta {\dot{f}} ^{\text{ Stage-2}}&= \DeltafdotFUTwo {~\text{Hz/s}}\\
    \Delta {\text{sky}} ^{\text{ Stage-2}}& \simeq \DeltaskyFUTwo ~\Delta {\text{sky}} ^{\text{ Stage-1}}
      \end{cases}
      \label{eq:FU3Box}
\end{equation}

\begin{figure}[h!tbp]
   \includegraphics[width=\columnwidth]{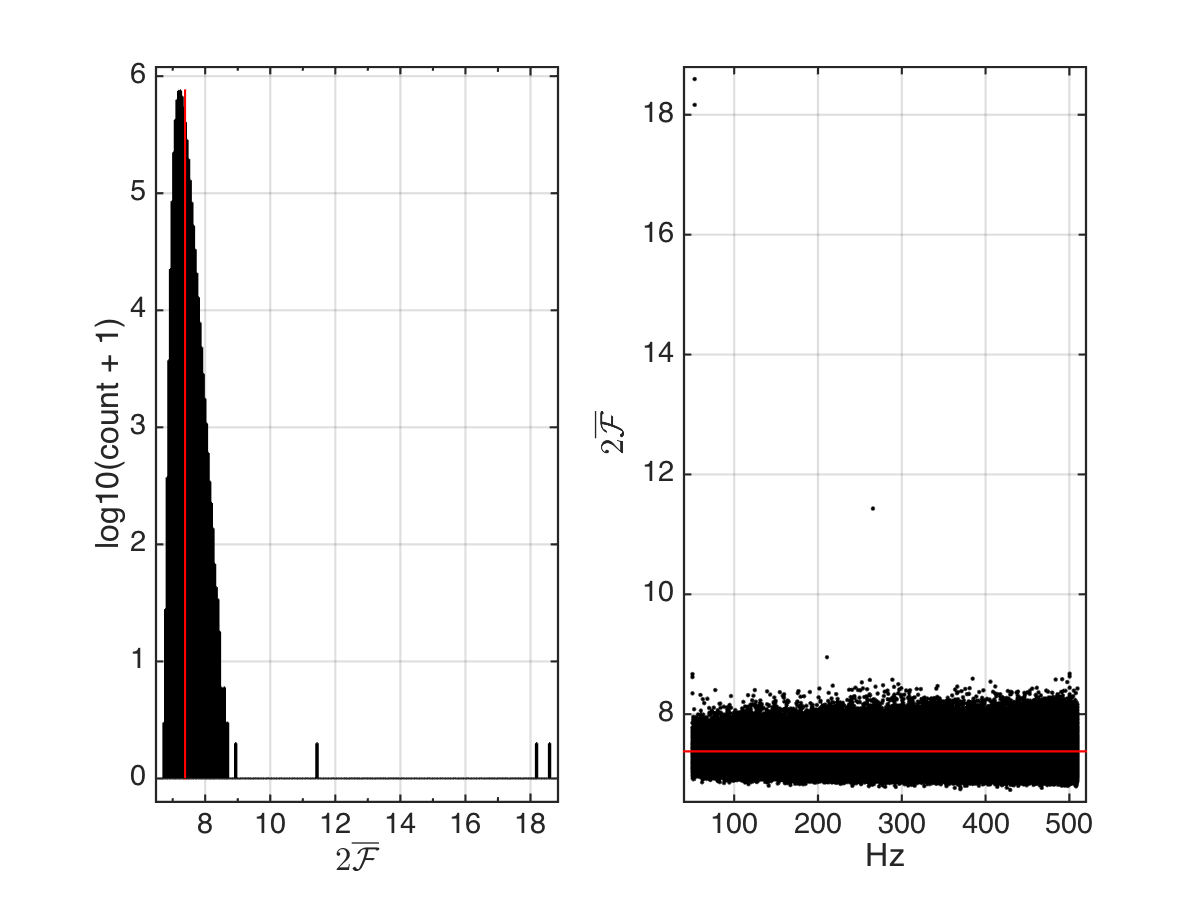}
\caption{Loudest from each of the stage-2 searches: the distribution of their detection statistic values $2\avF$ (left plot) and their distribution as a function of frequency (right plot). The red line marks $2\avF=\ThrFUTwo$ which is the threshold at and above which candidates are passed on to stage-3.}  
\label{fig:FU2Out}
\end{figure}

As done in stage 1 we record the most significant candidate from each search. The distribution is shown in Fig.\ref{fig:FU2Out}. In the next stage we follow-up the top 1.1 million candidates, corresponding to a threshold on $2\avF$ at \ThrFUTwo. This threshold has a \fdFUTwo false dismissal for signals at threshold and a \nrFUTwo noise rejection.


\begin{figure}[h!tbp]
   \includegraphics[width=\columnwidth]{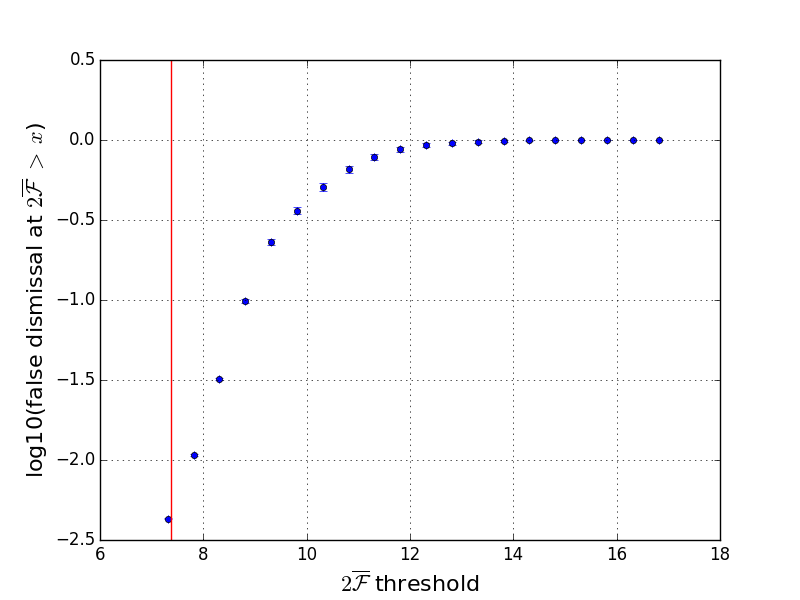}
\caption{Fraction of signals that are recovered with a detection statistic value larger than or equal to the threshold value after the stage-2 follow-up.}  
\label{fig:FU2FD}
\end{figure}

\subsection{Stage 3}
\label{subsec:stage3}

In this stage we search a volume of parameter space around each candidate defined by Eq.\ref{eq:FU3Box}.
As shown in Table \ref{tab:GridSpacings}, the coherent timebaseline is as long as that used in the previous stage but the grid spacings are finer. The search is divided among \TotWUsFUThree WUs each lasting about \WUCFUThree hours. 

The $>$ 99\% uncertainty regions for this search set-up for signals close to the detection threshold are 
\begin{equation}
  \begin{cases}
    \Delta f ^{\text{ Stage-3}}&= \DeltafFUThree {~\text{Hz}}\\
    \Delta {\dot{f}} ^{\text{ Stage-3}}&= \DeltafdotFUThree {~\text{Hz/s}}\\
    \Delta {\text{sky}} ^{\text{ Stage-3}}& \simeq \DeltaskyFUThree ~\Delta {\text{sky}} ^{\text{ Stage-2}}
      \end{cases}
      \label{eq:FU4Box}
\end{equation}

\begin{figure}[h!tbp]
   \includegraphics[width=\columnwidth]{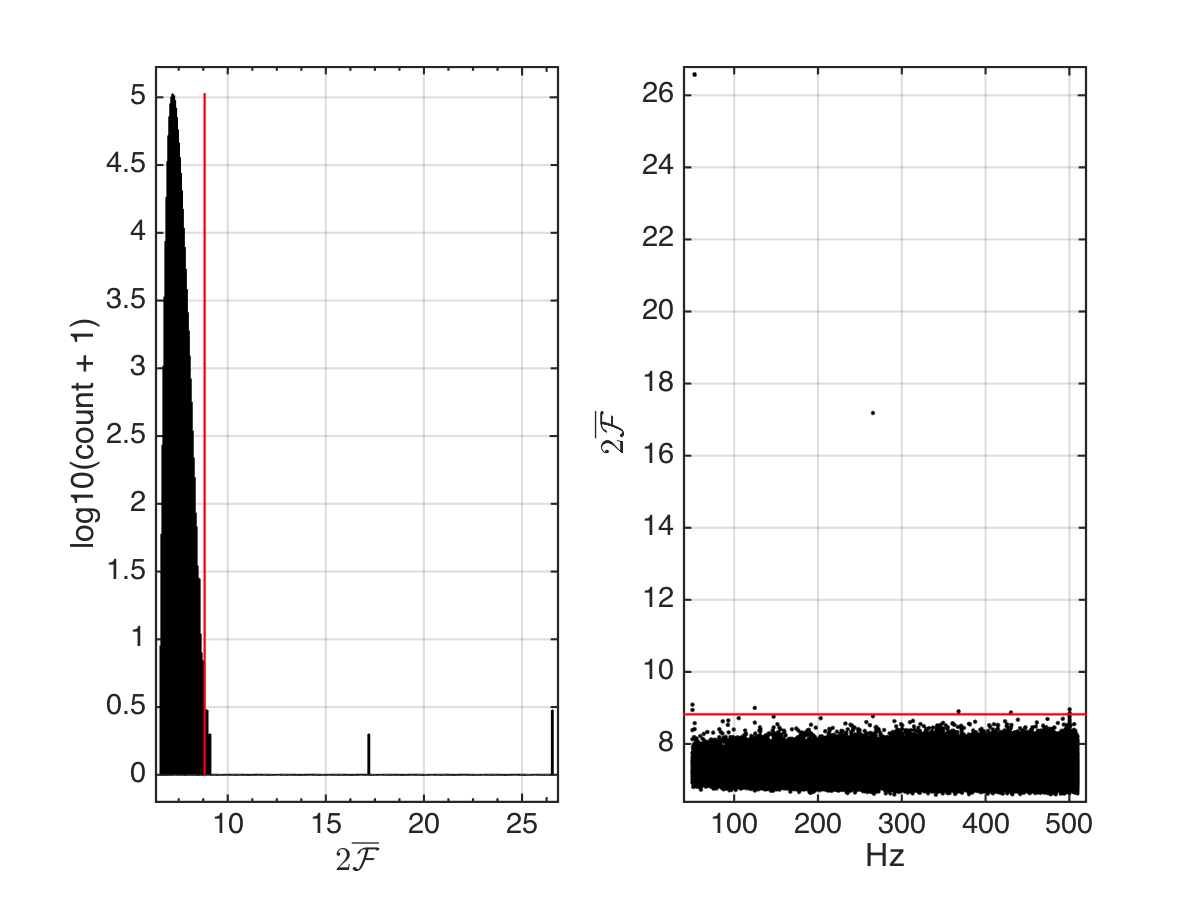}
\caption{Loudest from each of the stage-3 searches: the distribution of their detection statistic values $2\avF$ (left plot) and their distribution as a function of frequency (right plot). The red line marks $2\avF=8.82$ which is the threshold at and above which candidates are passed on to stage-4.}  
\label{fig:FU3Out}
\end{figure}

As done in previous stages we record the most significant candidate from each search. The distribution is shown in Fig.\ref{fig:FU3Out}. In the next stage we follow-up the top 10 candidates, corresponding to a threshold on $2\avF$ at \ThrFUThree. This threshold has a \fdFUThree false dismissal for signals at threshold and a \nrFUThree noise rejection.

\begin{figure}[h!tbp]
   \includegraphics[width=\columnwidth]{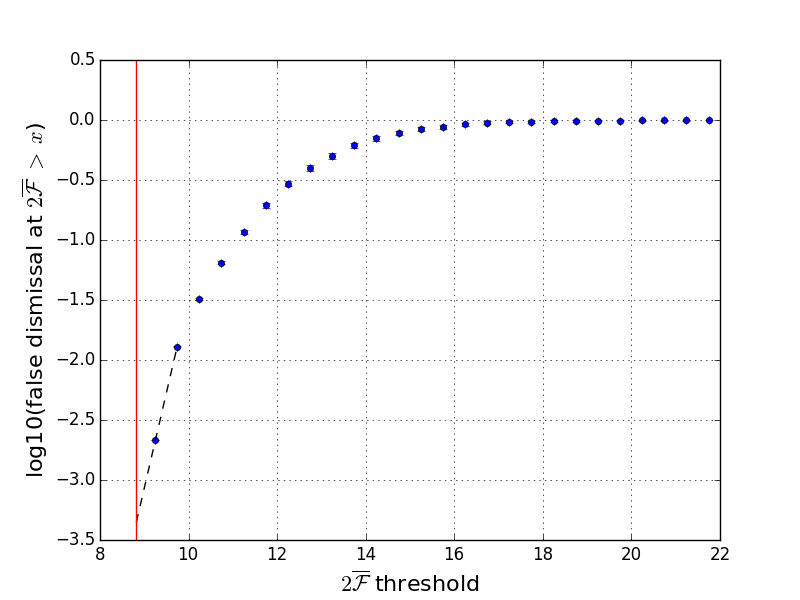}
\caption{Fraction of signals that are recovered with a detection statistic value larger than or equal to the threshold value (vertical line) after the stage-3 follow-up. The dashed line is a linear extrapolation based on the last two data points to guide the eye to the false dismissal value for signals at threshold. This line is a conservative estimate in the sense that it over estimates the false dismissal.}  
\label{fig:FU3FD}
\end{figure}

\subsection{Stage 4}
\label{subsec:stage4}

In this stage we search a volume of parameter space around each candidate defined by Eq.\ref{eq:FU4Box}.
The set-up of choice has a coherent time-baseline of \TcohFUFour hrs, twice as long as that used in stage-3, and the grid spacings shown in Table \ref{tab:GridSpacings}. The search has a relatively modest cost and is performed on the Atlas cluster: each follow-up lasts about {\WUCFUFour hours. The ranking statistic is $\OSNsc$ with a re-tuned $\cF=\cstarFUFour$. We consider the loudest candidate from each of the 10 follow-ups. 
In our Monte Carlo studies no signal candidate (out of 464 injections at threshold) is found more distant than:
\begin{equation}
  \begin{cases}
    \Delta f ^{\text{ Stage-4}}&= \DeltafFUFour {~\text{Hz}}\\
    \Delta {\dot{f}} ^{\text{ Stage-4}}&= \DeltafdotFUFour {~\text{Hz/s}}\\
    \Delta {\text{sky}} ^{\text{ Stage-4}}& \simeq \DeltaskyFUFour ~\Delta {\text{sky}} ^{\text{ Stage-3}}
      \end{cases}
      \label{eq:FU4Box}
\end{equation}
None of those injections has a $2\avF$ below \maxTwoFInjectionFUFour. Conservatively, we pick a threshold at \ThrFUFour. The Gaussian false alarm at $2\avF=$ \ThrFUFour is very low ($\approx$ \nrFUFour) and hence we do not expect any candidate  from random Gaussian noise fluctuations. 


Since we only follow-up 10 candidates we report our findings explicitly for each of them in Tab.\ref{tab:LastTenCandidates}. 

Candidates 3, 4 and 6 satisfy this condition but unfortunately they are ascribable to fake signals hardware-injected in the detector to test the detection pipelines. The search recovers all fake signals in the data with parameters within its search range and not absurdly loud\footnote{A fake signal was injected at about 108 Hz at such a high amplitude that it saturates the \EatH~ toplists across the entire sky. Upon visual inspection it is immediately obvious that the $f-\dot{f}$ morphology is that of a signal, albeit an unrealistically loud one. We categorised the associated band as disturbed because the data is corrupted by this loud injection and it is impossible to detect any real signal in its frequency neighbourhood.}. We note that candidates 3 and 4 come from the same fake signal. For a complete list of the fake signals present in the data -- see Table 6 of \cite{S6NineYoungSNRs}. In Table \ref{tab:HIParams} we show the signal parameters and report the distance with respect to the candidate parameter values. These distances are all within the Stage-4 uncertainties of \ref{eq:FU4Box}. We do not follow up these candidates any further because we know that they are associated with the hardware injections. 

The remaining candidates are below the threshold of  \ThrFUFour which is the minimum value of $2\avF$ that we demand candidates to pass before we inspect them further. 


\begin{figure}[h!tbp]
   \includegraphics[width=\columnwidth]{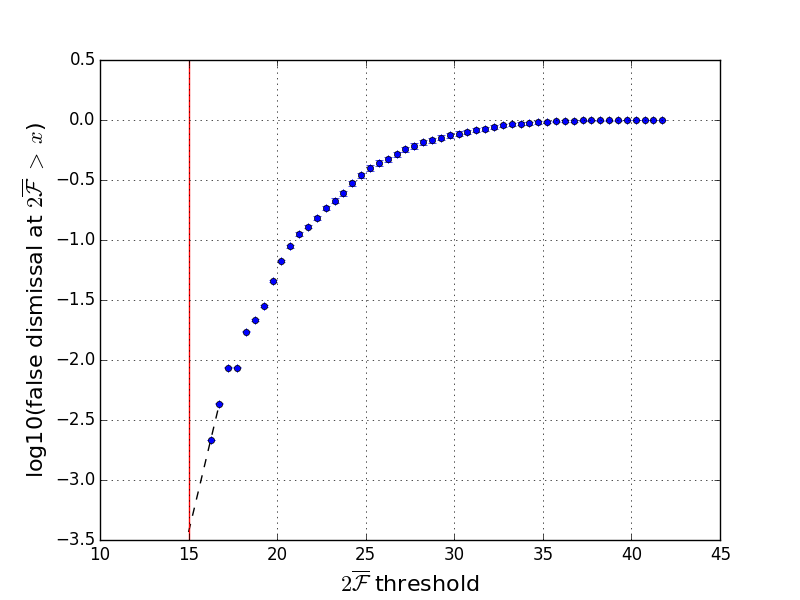}
\caption{Fraction of signals that are recovered with a detection statistic value larger than or equal to the threshold value (vertical line) after the stage-4 follow-up. The dashed line is a linear extrapolation based on the last two data points to guide the eye to the false dismissal value for signals at threshold. This line is a conservative estimate in the sense that it over estimates the false dismissal.}  
\label{fig:FU4FD}
\end{figure}

\begin{table*}[t]
\centering
\begin{tabular}{|c|c|c|c|c|c|c|c|}
\hline
ID & $f$  [Hz] & $\alpha$ [rad] & $\delta$ [rad] & ${\dot{f}}$ [Hz/s] & $2\avF$ & $2\avF_{\text{H1}}$ &  $2\avF_{\text{L1}}$\\
\hline
1&  \If & \Ialpha & \Idelta  & \Ifdot & \ITwoF  & \ITwoFH & \ITwoFL  \\
\hline
2 & \IIf &  \IIalpha& \IIdelta & \IIfdot & \IITwoF & \IITwoFH &  \IITwoFL \\
\hline
3& \IIIf & \IIIalpha & \IIIdelta & \IIIfdot & \IIITwoF & \IIITwoFH &  \IIITwoFL \\
\hline
4 &\IVf & \IValpha & \IVdelta & \IVfdot & \IVTwoF & \IVTwoFH &  \IVTwoFL \\
\hline
5& \Vf & \Valpha & \Vdelta & \Vfdot & \VTwoF & \VTwoFH &  \VTwoFL \\
\hline
6& \VIf & \VIalpha & \VIdelta & \VIfdot & \VITwoF & \VITwoFH &  \VITwoFL \\
\hline
7 &\VIIf &\VIIalpha & \VIIdelta & \VIIfdot & \VIITwoF & \VIITwoFH &  \VIITwoFL \\
\hline
8 & \VIIIf & \VIIIalpha & \VIIIdelta& \VIIIfdot & \VIIITwoF & \VIIITwoFH &  \VIIITwoFL \\
\hline
9 & \IXf & \IXalpha &\IXdelta& \IXfdot & \IXTwoF & \IXTwoFH &  \IXTwoFL \\
\hline
10 & \Xf & \Xalpha & \Xdelta & \Xfdot & \XTwoF & \XTwoFH &  \XTwoFL \\
\hline
\end{tabular}
\caption{Stage-4 results from each of the 10 follow-ups from the candidates surviving Stage-3. For illustration purposes in the last two columns we show the values of the average single-detector detection statistics. Typically for signals the single-detector values do not exceed the multi-detector $2\avF$.}
\label{tab:LastTenCandidates}
\end{table*}

\section{Results}
\label{sec:results}

\begin{table*}[t]
\centering
\begin{tabular}{|c|c|c|c|c|c|c|c|c|}
\hline
 ID & $f_{\text{s}}$  [Hz] &  $\alpha_{\text{s}} $ [rad] & $ \delta_{\text{s}} $ [rad] & $\dot{f_{\text{s}}}$  [Hz/s] & $ \Delta f$ [Hz] & $\Delta\alpha$ [rad] & $\Delta\delta$ [rad] & $\Delta {\dot{f}}$ [Hz/s] \\
\hline
3& \IIIinjf & \IIIinjalpha & \IIIinjdelta & \IIIinjfdot &\IIIDeltaf&\IIIDeltaalpha&\IIIDeltadelta&\IIIDeltafdot\\
\hline
 4& \IIIinjf & \IIIinjalpha &\IIIinjdelta &\IIIinjfdot &\IVDeltaf&\IVDeltaalpha&\IVDeltadelta&\IVDeltafdot\\
\hline
6 & \VIinjf& \VIinjalpha &\VIinjdelta&\VIinjfdot&\VIDeltaf&\VIDeltaalpha&\VIDeltadelta&\VIDeltafdot\\
\hline
\end{tabular}
\caption{Columns 2-5 show the parameters of the fake injected signal closest to the candidate whose ID identifies it in Table \ref{tab:LastTenCandidates}. Columns 6-9 display the distance between the candidates' and the signals' parameters (candidate parameter minus signal parameter). 
The reference time eference time (GPS s): \TrefFUfour}
\label{tab:HIParams}
\end{table*}

\begin{figure*}
   \includegraphics[width=0.8\textwidth]{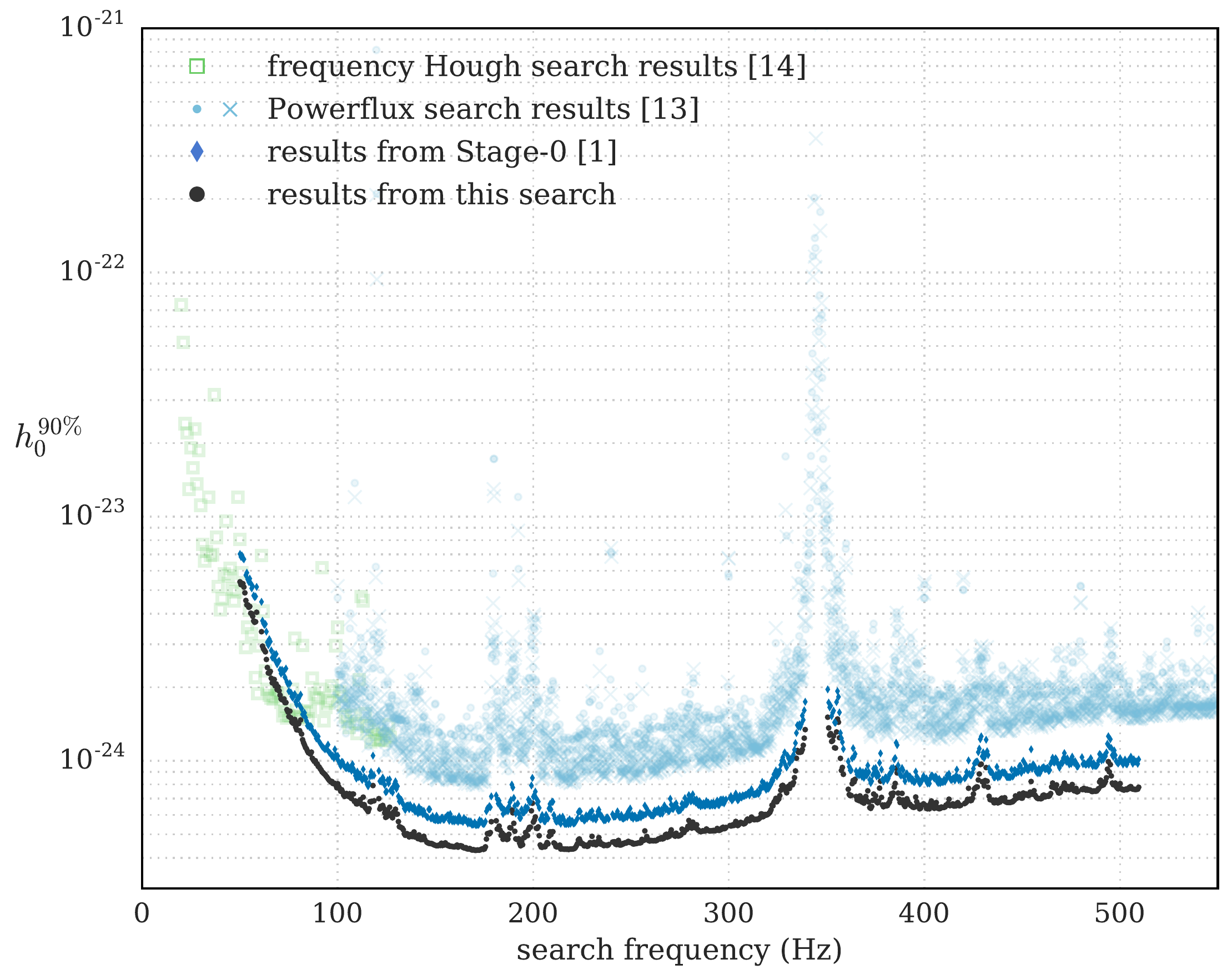}
\caption{90\% confidence upper limits on the gravitational wave amplitude of continuous gravitational wave signals with frequency in 0.5 Hz bands and with spindown values within the searched range. The lowest set of points (black circles) are the results of this search. For comparison we show the upper limits from only the stage-0 results \cite{S6BucketStage0}. These lie on the curve above the lowest one and are marked by dark blue diamonds. The results from a previous broad all-sky survey \cite{S6Powerflux} are the top curve (lighter circles and crosses) above 100 Hz. In the lower frequency we compare with a search on Virgo data contemporary to the LIGO S6 data \cite{AllSkyLowFreqVirgo}.}  
\label{fig:ULs}
\end{figure*}

The search did not reveal any continuous gravitational wave signal in the parameter volume that was searched. We hence set frequentist upper limits on the maximum gravitational wave amplitude consistent with this null result in $0.5$ Hz bands : $h_0^{90\%}$(f). $h_0^{90\%}$(f) is the GW amplitude such that 90\% of a population of signals with parameter values in our search range would have been detected by our search, i.e. would have survived the last $2\avF$ threshold at \ThrFUFour at stage-4. Since an actual full scale injection-and-recovery Monte Carlo for the entire set of follow-ups in every $0.5$ Hz band is prohibitive, in the same spirit as \cite{S6BucketStage0,S5GC1HF}, we perform such study in a limited set of trial bands. We pick \ULtrialBands. For each of these we determine the sensitivity depth of the search corresponding to the detection criterion stated above. As representative of the sensitivity depth ${{\mathcal{D}}}^{90\%}$ of this hierarchical search, we take the average of these depths, \AverageSD ~$[ {1/\sqrt{\text{Hz}}} ]$. Given the noise level of the data as a function of frequency, $S_h(f)$ , we then determine the 90\% upper limits as 
\begin{equation}
{ h{_{0}^{90\%}}(f) } = 
{
{\sqrt{S_h(f)}} \over { 
{{\mathcal{D}}}^{90\%} 
}
}.
\label{eq:ULfromSensDepth}
\end{equation}
Figure \ref{fig:ULs} shows these upper limits as a function of frequency. They are also presented in tabular form in the Appendix with the associated uncertainties which amount to \ULuncertainty, including calibration uncertainties. The most constraining upper limit is in the band between \MinFreqMin and \MinFreqMax Hz and it is \MinUL. At the upper end of the frequency range, around 510 Hz, the upper limit rises to  \ULHF.

The upper limits can be recast as exclusion regions in the signal frequency-ellipticity plane parameterised by the distance, for an isolated source emitting continuous gravitational waves due to its shape presenting an ellipticity $\epsilon$
\begin{equation}
\epsilon={|{I_{xx}-I_{yy}}|\over I_{zz}}
\label{eq:epsilon}
\end{equation}
where $I$ are the principal moments of inertia and the coordinate system is taken so that the $z$-axis is aligned with the spin axis of the star. Fig.\ref{fig:ULsEpsilon} shows these upper limits. Above 200 Hz we can exclude sources with ellipticities larger than $10^{-6}$ within 100 pc of Earth and above 400 Hz ellipticities above $4\times 10^{-7}$.

\begin{figure}[h!tbp]
   \includegraphics[width=\columnwidth]{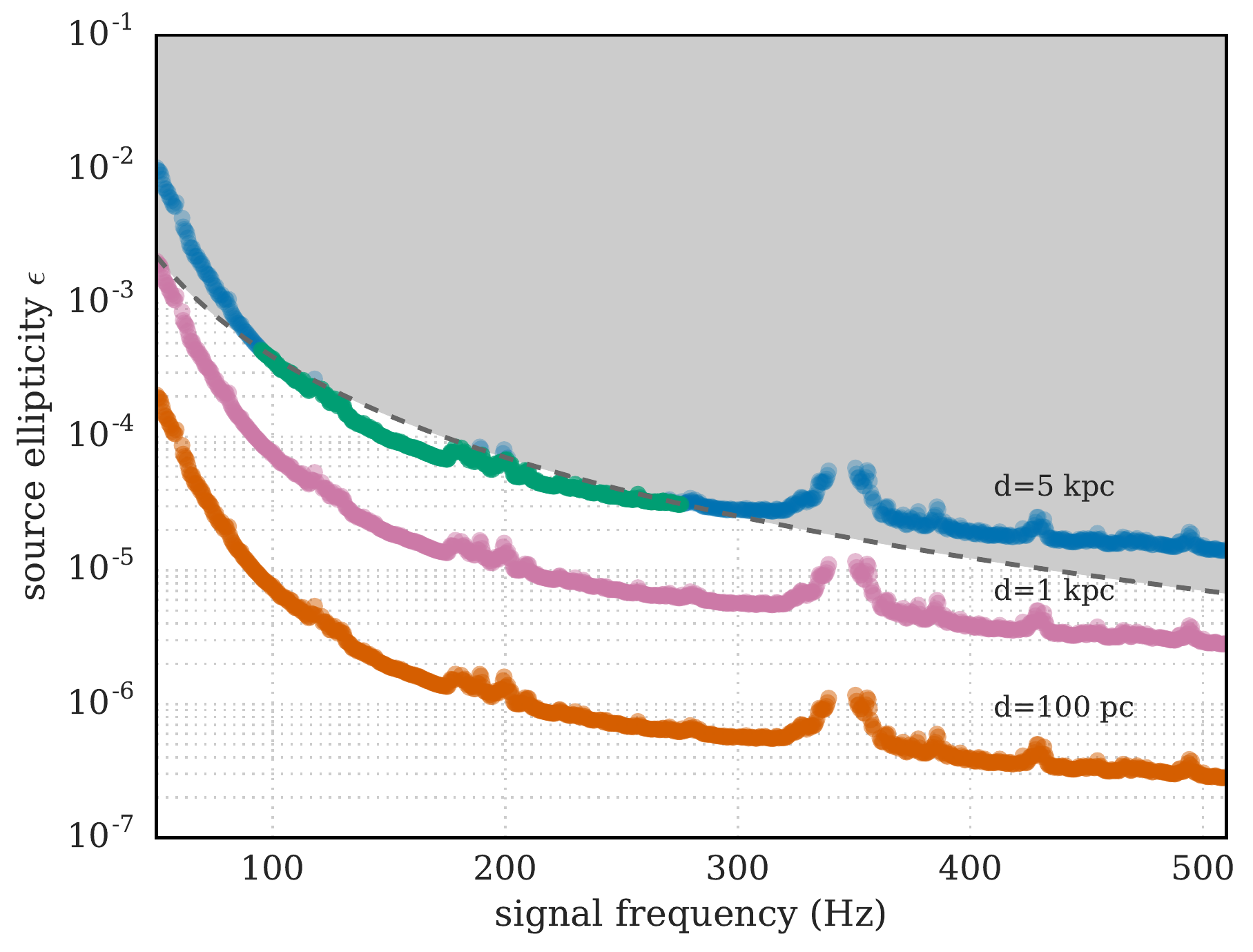}
\caption{Ellipticity $\epsilon$ of a source at a distance d emitting continuous gravitational waves that would have been detected by this search. The dashed line shows the spin-down ellipticity for the highest magnitude spindown parameter value searched: \maxfdot. The spin-down ellipticity is the ellipticity necessary for all the lost rotational kinetic energy to be emitted in gravitational waves. If we assume that the observed spin-down is all actual spin-down of the object, then no ellipticities could be possible above the dashed curve. In reality the observed and actual spindown could differ due to radial motion of the source. In this case the actual spin-down of the object may even be larger than the apparent one. In this case our search would be sensitive to objects with ellipticities above the dashed line. }  
\label{fig:ULsEpsilon}
\end{figure}


\section{Conclusions}
\label{sec:conclusions}

With a hierarchy of five semi-coherent searches at increasing coherent time baselines and resolutions in parameter space we searched over 16 million regions in few hundred Hz around the most sensitive frequencies of the LIGO detectors during the S6 science run. All stages but the very last ran on the \EatH~ distributed computing project lasting a few to several weeks. This is the first large-scale hierarchical search for gravitational wave signals ever performed. 

Having carried out this search proves that one can successfully perform deep follow-ups of marginal candidates and elevate their significance to the level necessary to be able to claim a detection. This paper proves that searches with thresholds at the level of the \EatH~ search described in \cite{Walsh:2016hyc} are possible; \cite{Walsh:2016hyc} demonstrates that they are the most sensitive and our observational results confirm this.

The sensitivity of broad surveys for continuous gravitational wave signals is computationally limited. For this reason we employ \EatH~ to deploy our searches. However, following up tens of millions of candidates is not just a matter of having the computational power. This paper illustrates how to perform and optimise the different stages, factoring in all the practical aspects of a real analysis.

None of the investigated candidates survived the five stages apart from those arising from the two fake signals injected in the detector for control purposes. These fake signals were recovered with the correct signal parameters. Candidate 6 comes from a hardware injection weak enough that no other search on this data set was ever able to detect it. This search recovers it well above the detection threshold. 

The gravitational wave amplitude upper limits that we set improve on existing ones \cite{S6BucketStage0} by about $30\%$. This corresponds to an increase in accessible space volume of $\simeq 2$. 

We excluded 10\% of the original data from this analysis because the stage-0 results had different statistical properties than the bulk of the results and the automated methods employed here which are necessary in order to deal with a large number of candidates, would not have yielded meaningful statistical results. We might go back to these excluded parameter space regions and attempt to extract information. This is a time-consuming process and the odds of finding a signal versus the odds of missing one by not analysing more sensitive data might well indicate that we shouldn't pursue this.

The optimal set-up for the various stages and the upper limits were determined at the expense of signal injection-and-recovery Monte Carlo studies. This is due to the fact that the implementation of a stack-slide search that we are using does not allow an analytical prediction of the sensitivity of a search with a given set-up (coherent segments and grid spacings). This major drawback will soon be overcome by a new implementation of stack-slide searches based on \cite{Wette:2016raf,Wette:2015lfa,Wette:2014tca,Wette:2013wza}. Such a search is being characterised and tuned at the time of writing and we hope to employ it in the context of our contributions to the LIGO Scientific Collaboration for searches on data from the next LIGO run (O2).

In principle we would like to carry out the entire hierarchy of stages on \EatH. For this to happen two aspects of the search presented here need to be automated: the visual inspection and the follow-up stages. The first is underway \cite{S6CasA}. The second will be significantly eased by the new stack-slide search that we alluded to, above.

\section{Acknowledgments}

Our sincere gratitude firstly goes to the \EatH $~$volunteers who have made these searches possible: thank you, thank you, thank you. Maria Alessandra Papa, Sin{\'e}ad Walsh, Bruce Allen and Xavier Siemens gratefully acknowledge the support from $\mathrm{NSF\;PHY}$ Grant 1104902. All the tuning and preparatory computational work for this search was carried out on the $\mathrm{ATLAS}$ super-computing cluster at the Max-Planck-Institut f{\"u}r Gravitationsphysik/ Leibniz Universit{\"a}t Hannover. We also acknowledge the Continuous Wave Group of the LIGO Scientific Collaboration for useful discussions and Graham Woan for reviewing the manuscript on behalf of the Collaboration. 

This document has been assigned LIGO Laboratory document number \texttt{LIGO-P1600213}.

\clearpage
\appendix
\onecolumngrid
\section{Tabular data}
\subsection{Upper limit values}
\label{A:ULs}
\onecolumngrid
\begin{longtable}{|c|c|c@{\hskip 0.1in}|c|c|c@{\hskip 0.1in}|c|c|c@{\hskip 0.1in}|c|c|}
\cline{1-2}\cline{4-5}\cline{7-8}\cline{10-11}
$\boldsymbol{f}$ \textbf{(in Hz)}& $\boldsymbol{h_{0}^{90\%}\times 10^{25}}$ & & $\boldsymbol{f}$ \textbf{(in Hz)} & $\boldsymbol{h_{0}^{90\%}\times 10^{25}}$ & & $\boldsymbol{f}$ \textbf{(in Hz)} & $\boldsymbol{h_{0}^{90\%}\times 10^{25}}$ & & $\boldsymbol{f}$ \textbf{(in Hz)} & $\boldsymbol{h_{0}^{90\%}\times 10^{25}}$\vspace{0.02in}\\
\cline{1-2}\cline{4-5}\cline{7-8}\cline{10-11}
\endhead
\centering
50.063 & 54.1 $\pm$ 10.8 & & 50.563 & 52.6 $\pm$ 10.5 & & 51.063 & 53.3 $\pm$ 10.7 & & 51.563 & 53.2 $\pm$ 10.6 \\
52.063 & 51.5 $\pm$ 10.3 & & 52.563 & 48.7 $\pm$ 9.7 & & 53.063 & 45.4 $\pm$ 9.1 & & 53.563 & 43.5 $\pm$ 8.7 \\
54.063 & 43.5 $\pm$ 8.7 & & 54.563 & 42.5 $\pm$ 8.5 & & 55.063 & 42.9 $\pm$ 8.6 & & 55.563 & 40.2 $\pm$ 8.0 \\
56.063 & 40.1 $\pm$ 8.0 & & 56.563 & 39.1 $\pm$ 7.8 & & 57.063 & 37.4 $\pm$ 7.5 & & 57.563 & 36.9 $\pm$ 7.4 \\
58.063 & 37.1 $\pm$ 7.4 & & 58.563 & 40.6 $\pm$ 8.1 & & 61.063 & 33.8 $\pm$ 6.8 & & 61.563 & 29.5 $\pm$ 5.9 \\
62.063 & 28.8 $\pm$ 5.8 & & 62.563 & 28.4 $\pm$ 5.7 & & 63.063 & 27.9 $\pm$ 5.6 & & 63.563 & 26.0 $\pm$ 5.2 \\
64.063 & 24.1 $\pm$ 4.8 & & 64.563 & 22.9 $\pm$ 4.6 & & 65.063 & 22.8 $\pm$ 4.6 & & 65.563 & 23.2 $\pm$ 4.6 \\
66.063 & 21.8 $\pm$ 4.4 & & 66.563 & 20.9 $\pm$ 4.2 & & 67.063 & 20.9 $\pm$ 4.2 & & 67.563 & 21.5 $\pm$ 4.3 \\
68.063 & 20.3 $\pm$ 4.1 & & 68.563 & 20.6 $\pm$ 4.1 & & 69.063 & 19.6 $\pm$ 3.9 & & 69.563 & 20.1 $\pm$ 4.0 \\
70.063 & 19.4 $\pm$ 3.9 & & 70.563 & 18.6 $\pm$ 3.7 & & 71.063 & 17.9 $\pm$ 3.6 & & 71.563 & 17.8 $\pm$ 3.6 \\
72.063 & 17.8 $\pm$ 3.6 & & 72.563 & 17.9 $\pm$ 3.6 & & 73.063 & 17.3 $\pm$ 3.5 & & 73.563 & 17.3 $\pm$ 3.5 \\
74.063 & 16.2 $\pm$ 3.2 & & 74.563 & 15.9 $\pm$ 3.2 & & 75.063 & 15.2 $\pm$ 3.0 & & 75.563 & 16.0 $\pm$ 3.2 \\
76.063 & 15.0 $\pm$ 3.0 & & 76.563 & 14.4 $\pm$ 2.9 & & 77.063 & 14.3 $\pm$ 2.9 & & 77.563 & 14.2 $\pm$ 2.8 \\
78.063 & 14.8 $\pm$ 3.0 & & 78.563 & 13.7 $\pm$ 2.7 & & 79.063 & 13.4 $\pm$ 2.7 & & 79.563 & 14.3 $\pm$ 2.9 \\
80.063 & 14.2 $\pm$ 2.8 & & 80.563 & 13.3 $\pm$ 2.7 & & 81.063 & 14.7 $\pm$ 2.9 & & 81.563 & 12.9 $\pm$ 2.6 \\
82.063 & 12.2 $\pm$ 2.4 & & 82.563 & 11.9 $\pm$ 2.4 & & 83.063 & 11.6 $\pm$ 2.3 & & 83.563 & 11.3 $\pm$ 2.3 \\
84.063 & 11.2 $\pm$ 2.2 & & 84.563 & 11.0 $\pm$ 2.2 & & 85.063 & 10.8 $\pm$ 2.2 & & 85.563 & 10.8 $\pm$ 2.2 \\
86.063 & 10.7 $\pm$ 2.1 & & 86.563 & 10.9 $\pm$ 2.2 & & 87.063 & 10.2 $\pm$ 2.0 & & 87.563 & 10.1 $\pm$ 2.0 \\
88.063 & 9.9 $\pm$ 2.0 & & 88.563 & 10.0 $\pm$ 2.0 & & 89.063 & 9.7 $\pm$ 1.9 & & 89.563 & 9.7 $\pm$ 1.9 \\
90.063 & 9.5 $\pm$ 1.9 & & 90.563 & 9.4 $\pm$ 1.9 & & 91.063 & 9.3 $\pm$ 1.9 & & 91.563 & 9.2 $\pm$ 1.8 \\
92.063 & 9.0 $\pm$ 1.8 & & 92.563 & 8.9 $\pm$ 1.8 & & 93.063 & 8.8 $\pm$ 1.8 & & 93.563 & 8.8 $\pm$ 1.8 \\
94.063 & 8.7 $\pm$ 1.7 & & 94.563 & 8.6 $\pm$ 1.7 & & 95.063 & 8.5 $\pm$ 1.7 & & 95.563 & 8.3 $\pm$ 1.7 \\
96.063 & 8.3 $\pm$ 1.7 & & 96.563 & 8.2 $\pm$ 1.6 & & 97.063 & 8.2 $\pm$ 1.6 & & 97.563 & 8.1 $\pm$ 1.6 \\
98.063 & 8.1 $\pm$ 1.6 & & 98.563 & 8.1 $\pm$ 1.6 & & 99.063 & 7.9 $\pm$ 1.6 & & 99.563 & 7.8 $\pm$ 1.6 \\
100.063 & 8.1 $\pm$ 1.6 & & 100.563 & 7.8 $\pm$ 1.6 & & 101.063 & 7.7 $\pm$ 1.5 & & 101.563 & 7.5 $\pm$ 1.5 \\
102.063 & 7.6 $\pm$ 1.5 & & 102.563 & 7.4 $\pm$ 1.5 & & 103.063 & 7.2 $\pm$ 1.4 & & 103.563 & 7.1 $\pm$ 1.4 \\
104.063 & 7.2 $\pm$ 1.4 & & 104.563 & 7.3 $\pm$ 1.5 & & 105.063 & 7.2 $\pm$ 1.4 & & 105.563 & 7.1 $\pm$ 1.4 \\
106.063 & 7.3 $\pm$ 1.5 & & 106.563 & 7.1 $\pm$ 1.4 & & 107.063 & 7.0 $\pm$ 1.4 & & 107.563 & 7.3 $\pm$ 1.5 \\
108.063 & 7.3 $\pm$ 1.5 & & 108.563 & 6.8 $\pm$ 1.4 & & 109.063 & 6.8 $\pm$ 1.4 & & 109.563 & 6.7 $\pm$ 1.3 \\
110.063 & 6.7 $\pm$ 1.3 & & 110.563 & 6.7 $\pm$ 1.3 & & 111.063 & 6.8 $\pm$ 1.4 & & 111.563 & 6.9 $\pm$ 1.4 \\
112.063 & 6.7 $\pm$ 1.3 & & 112.563 & 6.6 $\pm$ 1.3 & & 113.063 & 7.1 $\pm$ 1.4 & & 113.563 & 6.6 $\pm$ 1.3 \\
114.063 & 6.4 $\pm$ 1.3 & & 114.563 & 6.4 $\pm$ 1.3 & & 115.063 & 6.3 $\pm$ 1.3 & & 115.563 & 6.2 $\pm$ 1.2 \\
116.063 & 6.4 $\pm$ 1.3 & & 116.563 & 6.8 $\pm$ 1.4 & & 117.063 & 6.8 $\pm$ 1.4 & & 117.563 & 6.8 $\pm$ 1.4 \\
118.063 & 7.9 $\pm$ 1.6 & & 118.563 & 6.9 $\pm$ 1.4 & & 121.063 & 7.0 $\pm$ 1.4 & & 121.563 & 6.3 $\pm$ 1.3 \\
122.063 & 6.5 $\pm$ 1.3 & & 122.563 & 6.5 $\pm$ 1.3 & & 123.063 & 6.6 $\pm$ 1.3 & & 123.563 & 6.4 $\pm$ 1.3 \\
124.063 & 6.1 $\pm$ 1.2 & & 124.563 & 5.9 $\pm$ 1.2 & & 125.063 & 5.9 $\pm$ 1.2 & & 125.563 & 6.3 $\pm$ 1.3 \\
126.063 & 6.1 $\pm$ 1.2 & & 126.563 & 6.5 $\pm$ 1.3 & & 127.063 & 6.0 $\pm$ 1.2 & & 127.563 & 6.0 $\pm$ 1.2 \\
128.063 & 5.8 $\pm$ 1.2 & & 128.563 & 6.2 $\pm$ 1.2 & & 129.063 & 6.1 $\pm$ 1.2 & & 129.563 & 6.3 $\pm$ 1.3 \\
130.063 & 6.0 $\pm$ 1.2 & & 130.563 & 6.1 $\pm$ 1.2 & & 131.063 & 5.6 $\pm$ 1.1 & & 131.563 & 5.4 $\pm$ 1.1 \\
132.063 & 5.4 $\pm$ 1.1 & & 132.563 & 5.3 $\pm$ 1.1 & & 133.063 & 5.3 $\pm$ 1.1 & & 133.563 & 5.2 $\pm$ 1.0 \\
134.063 & 5.0 $\pm$ 1.0 & & 134.563 & 5.0 $\pm$ 1.0 & & 135.063 & 5.0 $\pm$ 1.0 & & 135.563 & 5.0 $\pm$ 1.0 \\
136.063 & 5.0 $\pm$ 1.0 & & 136.563 & 4.9 $\pm$ 1.0 & & 137.063 & 5.0 $\pm$ 1.0 & & 137.563 & 5.0 $\pm$ 1.0 \\
138.063 & 4.9 $\pm$ 1.0 & & 138.563 & 4.9 $\pm$ 1.0 & & 139.063 & 5.1 $\pm$ 1.0 & & 139.563 & 4.9 $\pm$ 1.0 \\
140.063 & 4.9 $\pm$ 1.0 & & 140.563 & 4.9 $\pm$ 1.0 & & 141.063 & 4.8 $\pm$ 1.0 & & 141.563 & 5.0 $\pm$ 1.0 \\
142.063 & 4.8 $\pm$ 1.0 & & 142.563 & 4.8 $\pm$ 1.0 & & 143.063 & 4.8 $\pm$ 1.0 & & 143.563 & 4.8 $\pm$ 1.0 \\
144.063 & 4.9 $\pm$ 1.0 & & 144.563 & 4.8 $\pm$ 1.0 & & 145.563 & 4.6 $\pm$ 0.9 & & 146.063 & 4.6 $\pm$ 0.9 \\
146.563 & 4.6 $\pm$ 0.9 & & 147.063 & 4.6 $\pm$ 0.9 & & 147.563 & 4.6 $\pm$ 0.9 & & 148.063 & 4.6 $\pm$ 0.9 \\
148.563 & 4.6 $\pm$ 0.9 & & 149.063 & 4.5 $\pm$ 0.9 & & 149.563 & 4.5 $\pm$ 0.9 & & 150.063 & 4.5 $\pm$ 0.9 \\
150.563 & 4.5 $\pm$ 0.9 & & 151.063 & 4.5 $\pm$ 0.9 & & 151.563 & 4.5 $\pm$ 0.9 & & 152.063 & 4.5 $\pm$ 0.9 \\
152.563 & 4.5 $\pm$ 0.9 & & 153.063 & 4.6 $\pm$ 0.9 & & 153.563 & 4.5 $\pm$ 0.9 & & 154.063 & 4.5 $\pm$ 0.9 \\
154.563 & 4.5 $\pm$ 0.9 & & 155.063 & 4.6 $\pm$ 0.9 & & 155.563 & 4.5 $\pm$ 0.9 & & 156.063 & 4.5 $\pm$ 0.9 \\
156.563 & 4.5 $\pm$ 0.9 & & 157.063 & 4.5 $\pm$ 0.9 & & 157.563 & 4.5 $\pm$ 0.9 & & 158.063 & 4.5 $\pm$ 0.9 \\
158.563 & 4.5 $\pm$ 0.9 & & 159.063 & 4.5 $\pm$ 0.9 & & 159.563 & 4.4 $\pm$ 0.9 & & 160.063 & 4.4 $\pm$ 0.9 \\
160.563 & 4.5 $\pm$ 0.9 & & 161.063 & 4.5 $\pm$ 0.9 & & 161.563 & 4.4 $\pm$ 0.9 & & 162.063 & 4.5 $\pm$ 0.9 \\
162.563 & 4.5 $\pm$ 0.9 & & 163.063 & 4.5 $\pm$ 0.9 & & 163.563 & 4.5 $\pm$ 0.9 & & 164.063 & 4.4 $\pm$ 0.9 \\
164.563 & 4.4 $\pm$ 0.9 & & 165.063 & 4.4 $\pm$ 0.9 & & 165.563 & 4.4 $\pm$ 0.9 & & 166.063 & 4.4 $\pm$ 0.9 \\
166.563 & 4.4 $\pm$ 0.9 & & 167.063 & 4.4 $\pm$ 0.9 & & 167.563 & 4.4 $\pm$ 0.9 & & 168.063 & 4.3 $\pm$ 0.9 \\
168.563 & 4.3 $\pm$ 0.9 & & 169.063 & 4.3 $\pm$ 0.9 & & 169.563 & 4.3 $\pm$ 0.9 & & 170.063 & 4.3 $\pm$ 0.9 \\
170.563 & 4.3 $\pm$ 0.9 & & 171.063 & 4.3 $\pm$ 0.9 & & 171.563 & 4.3 $\pm$ 0.9 & & 172.063 & 4.3 $\pm$ 0.9 \\
172.563 & 4.3 $\pm$ 0.9 & & 173.063 & 4.3 $\pm$ 0.9 & & 173.563 & 4.3 $\pm$ 0.9 & & 174.063 & 4.4 $\pm$ 0.9 \\
174.563 & 4.3 $\pm$ 0.9 & & 175.063 & 4.4 $\pm$ 0.9 & & 175.563 & 4.4 $\pm$ 0.9 & & 176.063 & 4.8 $\pm$ 1.0 \\
176.563 & 5.0 $\pm$ 1.0 & & 177.063 & 5.0 $\pm$ 1.0 & & 177.563 & 5.0 $\pm$ 1.0 & & 178.063 & 5.1 $\pm$ 1.0 \\
178.563 & 5.6 $\pm$ 1.1 & & 181.063 & 5.7 $\pm$ 1.1 & & 181.563 & 5.3 $\pm$ 1.1 & & 182.063 & 5.3 $\pm$ 1.1 \\
182.563 & 5.4 $\pm$ 1.1 & & 183.063 & 5.2 $\pm$ 1.0 & & 183.563 & 4.9 $\pm$ 1.0 & & 184.063 & 5.1 $\pm$ 1.0 \\
184.563 & 4.8 $\pm$ 1.0 & & 185.063 & 5.0 $\pm$ 1.0 & & 185.563 & 4.9 $\pm$ 1.0 & & 186.063 & 4.9 $\pm$ 1.0 \\
186.563 & 4.8 $\pm$ 1.0 & & 187.063 & 4.8 $\pm$ 1.0 & & 187.563 & 5.0 $\pm$ 1.0 & & 188.063 & 5.3 $\pm$ 1.1 \\
188.563 & 5.3 $\pm$ 1.1 & & 189.063 & 6.3 $\pm$ 1.3 & & 189.563 & 6.1 $\pm$ 1.2 & & 190.063 & 5.5 $\pm$ 1.1 \\
190.563 & 5.1 $\pm$ 1.0 & & 191.063 & 4.8 $\pm$ 1.0 & & 191.563 & 4.8 $\pm$ 1.0 & & 192.063 & 4.8 $\pm$ 1.0 \\
192.563 & 4.8 $\pm$ 1.0 & & 193.063 & 4.6 $\pm$ 0.9 & & 193.563 & 4.5 $\pm$ 0.9 & & 194.063 & 4.7 $\pm$ 0.9 \\
194.563 & 4.5 $\pm$ 0.9 & & 195.063 & 4.8 $\pm$ 1.0 & & 195.563 & 4.9 $\pm$ 1.0 & & 196.063 & 5.1 $\pm$ 1.0 \\
196.563 & 5.0 $\pm$ 1.0 & & 197.063 & 4.9 $\pm$ 1.0 & & 197.563 & 5.2 $\pm$ 1.0 & & 198.063 & 5.3 $\pm$ 1.1 \\
198.563 & 5.3 $\pm$ 1.1 & & 199.063 & 6.2 $\pm$ 1.2 & & 199.563 & 6.7 $\pm$ 1.3 & & 200.063 & 5.6 $\pm$ 1.1 \\
200.563 & 5.7 $\pm$ 1.1 & & 201.063 & 5.9 $\pm$ 1.2 & & 201.563 & 5.3 $\pm$ 1.1 & & 202.063 & 5.3 $\pm$ 1.1 \\
202.563 & 5.4 $\pm$ 1.1 & & 203.063 & 4.9 $\pm$ 1.0 & & 203.563 & 4.5 $\pm$ 0.9 & & 204.063 & 4.4 $\pm$ 0.9 \\
204.563 & 4.4 $\pm$ 0.9 & & 205.063 & 4.4 $\pm$ 0.9 & & 205.563 & 4.5 $\pm$ 0.9 & & 206.063 & 4.4 $\pm$ 0.9 \\
206.563 & 4.5 $\pm$ 0.9 & & 207.063 & 4.6 $\pm$ 0.9 & & 207.563 & 4.6 $\pm$ 0.9 & & 208.063 & 4.9 $\pm$ 1.0 \\
208.563 & 5.1 $\pm$ 1.0 & & 209.063 & 5.0 $\pm$ 1.0 & & 209.563 & 5.1 $\pm$ 1.0 & & 210.063 & 5.1 $\pm$ 1.0 \\
210.563 & 4.6 $\pm$ 0.9 & & 211.063 & 4.6 $\pm$ 0.9 & & 211.563 & 4.5 $\pm$ 0.9 & & 212.063 & 4.4 $\pm$ 0.9 \\
212.563 & 4.4 $\pm$ 0.9 & & 213.063 & 4.5 $\pm$ 0.9 & & 213.563 & 4.5 $\pm$ 0.9 & & 214.063 & 4.3 $\pm$ 0.9 \\
214.563 & 4.4 $\pm$ 0.9 & & 215.063 & 4.4 $\pm$ 0.9 & & 215.563 & 4.3 $\pm$ 0.9 & & 216.063 & 4.3 $\pm$ 0.9 \\
216.563 & 4.3 $\pm$ 0.9 & & 217.063 & 4.3 $\pm$ 0.9 & & 217.563 & 4.3 $\pm$ 0.9 & & 218.063 & 4.3 $\pm$ 0.9 \\
218.563 & 4.3 $\pm$ 0.9 & & 219.063 & 4.4 $\pm$ 0.9 & & 219.563 & 4.3 $\pm$ 0.9 & & 220.063 & 4.4 $\pm$ 0.9 \\
220.563 & 4.4 $\pm$ 0.9 & & 221.063 & 4.4 $\pm$ 0.9 & & 221.563 & 4.4 $\pm$ 0.9 & & 222.063 & 4.5 $\pm$ 0.9 \\
222.563 & 4.6 $\pm$ 0.9 & & 223.063 & 4.7 $\pm$ 0.9 & & 223.563 & 4.8 $\pm$ 1.0 & & 224.063 & 4.7 $\pm$ 0.9 \\
224.563 & 4.6 $\pm$ 0.9 & & 225.063 & 4.6 $\pm$ 0.9 & & 225.563 & 4.6 $\pm$ 0.9 & & 226.063 & 4.5 $\pm$ 0.9 \\
226.563 & 4.5 $\pm$ 0.9 & & 227.063 & 4.5 $\pm$ 0.9 & & 227.563 & 4.5 $\pm$ 0.9 & & 228.063 & 4.5 $\pm$ 0.9 \\
228.563 & 4.6 $\pm$ 0.9 & & 229.063 & 4.6 $\pm$ 0.9 & & 229.563 & 4.6 $\pm$ 0.9 & & 230.063 & 4.9 $\pm$ 1.0 \\
230.563 & 4.6 $\pm$ 0.9 & & 231.063 & 4.6 $\pm$ 0.9 & & 231.563 & 4.6 $\pm$ 0.9 & & 232.063 & 4.5 $\pm$ 0.9 \\
232.563 & 4.6 $\pm$ 0.9 & & 233.063 & 4.7 $\pm$ 0.9 & & 233.563 & 4.8 $\pm$ 1.0 & & 234.063 & 4.7 $\pm$ 0.9 \\
234.563 & 4.6 $\pm$ 0.9 & & 235.063 & 4.6 $\pm$ 0.9 & & 235.563 & 4.6 $\pm$ 0.9 & & 236.063 & 4.5 $\pm$ 0.9 \\
236.563 & 4.5 $\pm$ 0.9 & & 237.063 & 4.5 $\pm$ 0.9 & & 237.563 & 4.5 $\pm$ 0.9 & & 238.063 & 4.5 $\pm$ 0.9 \\
238.563 & 4.5 $\pm$ 0.9 & & 240.563 & 4.6 $\pm$ 0.9 & & 241.063 & 4.6 $\pm$ 0.9 & & 241.563 & 4.7 $\pm$ 0.9 \\
242.063 & 4.6 $\pm$ 0.9 & & 242.563 & 4.5 $\pm$ 0.9 & & 243.063 & 4.7 $\pm$ 0.9 & & 243.563 & 4.7 $\pm$ 0.9 \\
244.063 & 4.5 $\pm$ 0.9 & & 244.563 & 4.5 $\pm$ 0.9 & & 245.063 & 4.5 $\pm$ 0.9 & & 245.563 & 4.6 $\pm$ 0.9 \\
246.063 & 4.6 $\pm$ 0.9 & & 246.563 & 4.6 $\pm$ 0.9 & & 247.063 & 4.6 $\pm$ 0.9 & & 247.563 & 4.6 $\pm$ 0.9 \\
248.063 & 4.6 $\pm$ 0.9 & & 248.563 & 4.7 $\pm$ 0.9 & & 249.063 & 4.7 $\pm$ 0.9 & & 249.563 & 4.6 $\pm$ 0.9 \\
250.063 & 4.6 $\pm$ 0.9 & & 250.563 & 4.6 $\pm$ 0.9 & & 251.063 & 4.6 $\pm$ 0.9 & & 251.563 & 4.6 $\pm$ 0.9 \\
252.063 & 4.6 $\pm$ 0.9 & & 252.563 & 4.6 $\pm$ 0.9 & & 253.063 & 4.6 $\pm$ 0.9 & & 253.563 & 4.6 $\pm$ 0.9 \\
254.063 & 4.6 $\pm$ 0.9 & & 254.563 & 4.6 $\pm$ 0.9 & & 255.063 & 4.6 $\pm$ 0.9 & & 255.563 & 4.8 $\pm$ 1.0 \\
256.063 & 4.7 $\pm$ 0.9 & & 256.563 & 4.7 $\pm$ 0.9 & & 257.063 & 5.2 $\pm$ 1.0 & & 257.563 & 4.8 $\pm$ 1.0 \\
258.063 & 4.9 $\pm$ 1.0 & & 258.563 & 4.8 $\pm$ 1.0 & & 259.063 & 4.7 $\pm$ 0.9 & & 259.563 & 4.7 $\pm$ 0.9 \\
260.063 & 4.7 $\pm$ 0.9 & & 260.563 & 4.7 $\pm$ 0.9 & & 261.063 & 4.7 $\pm$ 0.9 & & 261.563 & 4.7 $\pm$ 0.9 \\
262.063 & 4.7 $\pm$ 0.9 & & 262.563 & 4.7 $\pm$ 0.9 & & 263.063 & 4.7 $\pm$ 0.9 & & 263.563 & 4.7 $\pm$ 0.9 \\
264.063 & 4.8 $\pm$ 1.0 & & 264.563 & 4.8 $\pm$ 1.0 & & 265.063 & 4.8 $\pm$ 1.0 & & 265.563 & 4.8 $\pm$ 1.0 \\
266.063 & 4.8 $\pm$ 1.0 & & 266.563 & 4.8 $\pm$ 1.0 & & 267.063 & 4.8 $\pm$ 1.0 & & 267.563 & 5.0 $\pm$ 1.0 \\
268.063 & 5.0 $\pm$ 1.0 & & 268.563 & 4.9 $\pm$ 1.0 & & 269.063 & 4.9 $\pm$ 1.0 & & 269.563 & 4.9 $\pm$ 1.0 \\
270.063 & 5.1 $\pm$ 1.0 & & 270.563 & 5.2 $\pm$ 1.0 & & 271.063 & 5.0 $\pm$ 1.0 & & 271.563 & 5.0 $\pm$ 1.0 \\
272.063 & 4.9 $\pm$ 1.0 & & 272.563 & 4.9 $\pm$ 1.0 & & 273.063 & 5.0 $\pm$ 1.0 & & 273.563 & 5.0 $\pm$ 1.0 \\
274.063 & 4.9 $\pm$ 1.0 & & 274.563 & 4.9 $\pm$ 1.0 & & 275.063 & 4.9 $\pm$ 1.0 & & 275.563 & 5.0 $\pm$ 1.0 \\
276.063 & 5.3 $\pm$ 1.1 & & 276.563 & 5.1 $\pm$ 1.0 & & 277.063 & 5.1 $\pm$ 1.0 & & 277.563 & 5.2 $\pm$ 1.0 \\
278.063 & 5.2 $\pm$ 1.0 & & 278.563 & 5.2 $\pm$ 1.0 & & 279.063 & 5.4 $\pm$ 1.1 & & 279.563 & 5.7 $\pm$ 1.1 \\
280.063 & 5.5 $\pm$ 1.1 & & 280.563 & 5.4 $\pm$ 1.1 & & 281.063 & 5.3 $\pm$ 1.1 & & 281.563 & 5.6 $\pm$ 1.1 \\
282.063 & 5.4 $\pm$ 1.1 & & 282.563 & 5.3 $\pm$ 1.1 & & 283.063 & 5.3 $\pm$ 1.1 & & 283.563 & 5.5 $\pm$ 1.1 \\
284.063 & 5.2 $\pm$ 1.0 & & 284.563 & 5.2 $\pm$ 1.0 & & 285.063 & 5.2 $\pm$ 1.0 & & 285.563 & 5.1 $\pm$ 1.0 \\
286.063 & 5.1 $\pm$ 1.0 & & 286.563 & 5.2 $\pm$ 1.0 & & 287.063 & 5.2 $\pm$ 1.0 & & 287.563 & 5.2 $\pm$ 1.0 \\
288.063 & 5.2 $\pm$ 1.0 & & 288.563 & 5.3 $\pm$ 1.1 & & 289.063 & 5.2 $\pm$ 1.0 & & 289.563 & 5.2 $\pm$ 1.0 \\
290.063 & 5.2 $\pm$ 1.0 & & 290.563 & 5.2 $\pm$ 1.0 & & 291.063 & 5.2 $\pm$ 1.0 & & 291.563 & 5.2 $\pm$ 1.0 \\
292.063 & 5.2 $\pm$ 1.0 & & 292.563 & 5.2 $\pm$ 1.0 & & 293.063 & 5.2 $\pm$ 1.0 & & 293.563 & 5.2 $\pm$ 1.0 \\
294.063 & 5.3 $\pm$ 1.1 & & 294.563 & 5.2 $\pm$ 1.0 & & 295.063 & 5.2 $\pm$ 1.0 & & 295.563 & 5.2 $\pm$ 1.0 \\
296.063 & 5.2 $\pm$ 1.0 & & 296.563 & 5.3 $\pm$ 1.1 & & 297.063 & 5.3 $\pm$ 1.1 & & 297.563 & 5.3 $\pm$ 1.1 \\
298.063 & 5.3 $\pm$ 1.1 & & 298.563 & 5.3 $\pm$ 1.1 & & 300.563 & 5.4 $\pm$ 1.1 & & 301.063 & 5.4 $\pm$ 1.1 \\
301.563 & 5.4 $\pm$ 1.1 & & 302.063 & 5.5 $\pm$ 1.1 & & 302.563 & 5.4 $\pm$ 1.1 & & 303.063 & 5.5 $\pm$ 1.1 \\
303.563 & 5.6 $\pm$ 1.1 & & 304.063 & 5.5 $\pm$ 1.1 & & 304.563 & 5.4 $\pm$ 1.1 & & 305.063 & 5.5 $\pm$ 1.1 \\
305.563 & 5.5 $\pm$ 1.1 & & 306.063 & 5.6 $\pm$ 1.1 & & 306.563 & 5.6 $\pm$ 1.1 & & 307.063 & 5.5 $\pm$ 1.1 \\
307.563 & 5.5 $\pm$ 1.1 & & 308.063 & 5.5 $\pm$ 1.1 & & 308.563 & 5.6 $\pm$ 1.1 & & 309.063 & 5.7 $\pm$ 1.1 \\
309.563 & 5.8 $\pm$ 1.2 & & 310.063 & 5.7 $\pm$ 1.1 & & 310.563 & 5.7 $\pm$ 1.1 & & 311.063 & 5.7 $\pm$ 1.1 \\
311.563 & 5.9 $\pm$ 1.2 & & 312.063 & 5.8 $\pm$ 1.2 & & 312.563 & 5.7 $\pm$ 1.1 & & 313.063 & 5.7 $\pm$ 1.1 \\
313.563 & 5.8 $\pm$ 1.2 & & 314.063 & 5.8 $\pm$ 1.2 & & 314.563 & 5.7 $\pm$ 1.1 & & 315.063 & 5.8 $\pm$ 1.2 \\
315.563 & 5.8 $\pm$ 1.2 & & 316.063 & 5.9 $\pm$ 1.2 & & 316.563 & 6.1 $\pm$ 1.2 & & 317.063 & 6.0 $\pm$ 1.2 \\
317.563 & 5.9 $\pm$ 1.2 & & 318.063 & 6.0 $\pm$ 1.2 & & 318.563 & 6.0 $\pm$ 1.2 & & 319.063 & 6.0 $\pm$ 1.2 \\
319.563 & 6.0 $\pm$ 1.2 & & 320.063 & 6.0 $\pm$ 1.2 & & 320.563 & 6.1 $\pm$ 1.2 & & 321.063 & 6.2 $\pm$ 1.2 \\
321.563 & 6.3 $\pm$ 1.3 & & 322.063 & 6.6 $\pm$ 1.3 & & 322.563 & 6.5 $\pm$ 1.3 & & 323.063 & 6.8 $\pm$ 1.4 \\
323.563 & 6.9 $\pm$ 1.4 & & 324.063 & 7.0 $\pm$ 1.4 & & 324.563 & 6.8 $\pm$ 1.4 & & 325.063 & 6.9 $\pm$ 1.4 \\
325.563 & 7.0 $\pm$ 1.4 & & 326.063 & 7.2 $\pm$ 1.4 & & 326.563 & 7.6 $\pm$ 1.5 & & 327.063 & 7.9 $\pm$ 1.6 \\
327.563 & 7.9 $\pm$ 1.6 & & 328.063 & 7.8 $\pm$ 1.6 & & 328.563 & 7.7 $\pm$ 1.5 & & 329.063 & 7.5 $\pm$ 1.5 \\
329.563 & 7.4 $\pm$ 1.5 & & 330.063 & 7.7 $\pm$ 1.5 & & 330.563 & 7.9 $\pm$ 1.6 & & 331.063 & 7.7 $\pm$ 1.5 \\
331.563 & 8.0 $\pm$ 1.6 & & 332.063 & 8.0 $\pm$ 1.6 & & 332.563 & 8.0 $\pm$ 1.6 & & 333.063 & 8.1 $\pm$ 1.6 \\
333.563 & 8.5 $\pm$ 1.7 & & 334.063 & 9.1 $\pm$ 1.8 & & 334.563 & 10.2 $\pm$ 2.0 & & 335.063 & 11.0 $\pm$ 2.2 \\
335.563 & 10.8 $\pm$ 2.2 & & 336.063 & 10.8 $\pm$ 2.2 & & 336.563 & 10.8 $\pm$ 2.2 & & 337.063 & 10.9 $\pm$ 2.2 \\
337.563 & 11.1 $\pm$ 2.2 & & 338.063 & 11.6 $\pm$ 2.3 & & 338.563 & 12.4 $\pm$ 2.5 & & 339.063 & 13.4 $\pm$ 2.7 \\
350.563 & 15.1 $\pm$ 3.0 & & 351.063 & 13.6 $\pm$ 2.7 & & 351.563 & 12.7 $\pm$ 2.5 & & 352.063 & 12.9 $\pm$ 2.6 \\
352.563 & 12.0 $\pm$ 2.4 & & 353.063 & 12.1 $\pm$ 2.4 & & 353.563 & 12.6 $\pm$ 2.5 & & 354.063 & 11.3 $\pm$ 2.3 \\
354.563 & 11.3 $\pm$ 2.3 & & 355.063 & 13.1 $\pm$ 2.6 & & 355.563 & 14.8 $\pm$ 3.0 & & 356.063 & 14.4 $\pm$ 2.9 \\
356.563 & 12.4 $\pm$ 2.5 & & 357.063 & 10.2 $\pm$ 2.0 & & 357.563 & 9.1 $\pm$ 1.8 & & 358.063 & 9.4 $\pm$ 1.9 \\
358.563 & 8.8 $\pm$ 1.8 & & 361.063 & 7.6 $\pm$ 1.5 & & 361.563 & 7.3 $\pm$ 1.5 & & 362.063 & 7.2 $\pm$ 1.4 \\
362.563 & 7.2 $\pm$ 1.4 & & 363.063 & 8.1 $\pm$ 1.6 & & 363.563 & 8.3 $\pm$ 1.7 & & 364.063 & 8.2 $\pm$ 1.6 \\
364.563 & 8.4 $\pm$ 1.7 & & 365.063 & 7.1 $\pm$ 1.4 & & 365.563 & 6.9 $\pm$ 1.4 & & 366.063 & 7.0 $\pm$ 1.4 \\
366.563 & 6.9 $\pm$ 1.4 & & 367.063 & 7.2 $\pm$ 1.4 & & 367.563 & 7.1 $\pm$ 1.4 & & 368.063 & 6.8 $\pm$ 1.4 \\
368.563 & 6.9 $\pm$ 1.4 & & 369.063 & 6.7 $\pm$ 1.3 & & 369.563 & 7.0 $\pm$ 1.4 & & 370.063 & 7.1 $\pm$ 1.4 \\
370.563 & 6.9 $\pm$ 1.4 & & 371.063 & 7.5 $\pm$ 1.5 & & 371.563 & 6.8 $\pm$ 1.4 & & 372.063 & 6.4 $\pm$ 1.3 \\
372.563 & 6.4 $\pm$ 1.3 & & 373.063 & 6.5 $\pm$ 1.3 & & 373.563 & 6.9 $\pm$ 1.4 & & 374.063 & 7.3 $\pm$ 1.5 \\
374.563 & 6.9 $\pm$ 1.4 & & 375.063 & 7.2 $\pm$ 1.4 & & 375.563 & 6.8 $\pm$ 1.4 & & 376.063 & 6.7 $\pm$ 1.3 \\
376.563 & 6.8 $\pm$ 1.4 & & 377.063 & 7.7 $\pm$ 1.5 & & 377.563 & 8.3 $\pm$ 1.7 & & 378.063 & 7.1 $\pm$ 1.4 \\
378.563 & 6.6 $\pm$ 1.3 & & 379.063 & 6.5 $\pm$ 1.3 & & 379.563 & 6.6 $\pm$ 1.3 & & 380.063 & 6.6 $\pm$ 1.3 \\
380.563 & 6.5 $\pm$ 1.3 & & 381.063 & 6.6 $\pm$ 1.3 & & 381.563 & 6.7 $\pm$ 1.3 & & 382.063 & 6.8 $\pm$ 1.4 \\
382.563 & 7.0 $\pm$ 1.4 & & 383.063 & 7.3 $\pm$ 1.5 & & 383.563 & 7.2 $\pm$ 1.4 & & 384.063 & 7.4 $\pm$ 1.5 \\
384.563 & 7.8 $\pm$ 1.6 & & 385.063 & 8.1 $\pm$ 1.6 & & 385.563 & 9.3 $\pm$ 1.9 & & 386.063 & 8.9 $\pm$ 1.8 \\
386.563 & 7.4 $\pm$ 1.5 & & 387.063 & 7.0 $\pm$ 1.4 & & 387.563 & 6.8 $\pm$ 1.4 & & 388.063 & 6.9 $\pm$ 1.4 \\
388.563 & 7.4 $\pm$ 1.5 & & 389.063 & 6.9 $\pm$ 1.4 & & 389.563 & 6.6 $\pm$ 1.3 & & 390.063 & 6.5 $\pm$ 1.3 \\
390.563 & 6.8 $\pm$ 1.4 & & 391.063 & 7.0 $\pm$ 1.4 & & 391.563 & 6.8 $\pm$ 1.4 & & 392.063 & 6.6 $\pm$ 1.3 \\
392.563 & 6.6 $\pm$ 1.3 & & 393.063 & 6.6 $\pm$ 1.3 & & 393.563 & 6.5 $\pm$ 1.3 & & 394.063 & 6.5 $\pm$ 1.3 \\
394.563 & 6.4 $\pm$ 1.3 & & 395.063 & 6.5 $\pm$ 1.3 & & 395.563 & 7.1 $\pm$ 1.4 & & 396.063 & 6.8 $\pm$ 1.4 \\
396.563 & 6.6 $\pm$ 1.3 & & 397.063 & 6.6 $\pm$ 1.3 & & 397.563 & 6.4 $\pm$ 1.3 & & 398.063 & 6.4 $\pm$ 1.3 \\
398.563 & 6.6 $\pm$ 1.3 & & 399.063 & 6.6 $\pm$ 1.3 & & 399.563 & 6.7 $\pm$ 1.3 & & 400.563 & 6.6 $\pm$ 1.3 \\
401.063 & 6.4 $\pm$ 1.3 & & 401.563 & 6.4 $\pm$ 1.3 & & 402.063 & 6.4 $\pm$ 1.3 & & 402.563 & 6.4 $\pm$ 1.3 \\
403.063 & 6.7 $\pm$ 1.3 & & 403.563 & 6.8 $\pm$ 1.4 & & 404.063 & 6.7 $\pm$ 1.3 & & 404.563 & 6.5 $\pm$ 1.3 \\
405.063 & 6.4 $\pm$ 1.3 & & 405.563 & 6.6 $\pm$ 1.3 & & 406.063 & 6.7 $\pm$ 1.3 & & 406.563 & 6.5 $\pm$ 1.3 \\
407.063 & 6.4 $\pm$ 1.3 & & 407.563 & 6.4 $\pm$ 1.3 & & 408.063 & 6.4 $\pm$ 1.3 & & 408.563 & 6.5 $\pm$ 1.3 \\
409.063 & 6.5 $\pm$ 1.3 & & 409.563 & 6.4 $\pm$ 1.3 & & 410.063 & 6.4 $\pm$ 1.3 & & 410.563 & 6.5 $\pm$ 1.3 \\
411.063 & 6.6 $\pm$ 1.3 & & 411.563 & 6.6 $\pm$ 1.3 & & 412.063 & 6.7 $\pm$ 1.3 & & 412.563 & 7.0 $\pm$ 1.4 \\
413.063 & 6.6 $\pm$ 1.3 & & 413.563 & 6.6 $\pm$ 1.3 & & 414.063 & 6.6 $\pm$ 1.3 & & 414.563 & 6.7 $\pm$ 1.3 \\
415.063 & 6.5 $\pm$ 1.3 & & 415.563 & 6.5 $\pm$ 1.3 & & 416.063 & 6.5 $\pm$ 1.3 & & 416.563 & 6.7 $\pm$ 1.3 \\
417.063 & 6.7 $\pm$ 1.3 & & 417.563 & 6.6 $\pm$ 1.3 & & 418.063 & 6.5 $\pm$ 1.3 & & 418.563 & 6.6 $\pm$ 1.3 \\
420.563 & 6.7 $\pm$ 1.3 & & 421.063 & 6.7 $\pm$ 1.3 & & 421.563 & 6.8 $\pm$ 1.4 & & 422.063 & 7.0 $\pm$ 1.4 \\
422.563 & 7.7 $\pm$ 1.5 & & 423.063 & 7.0 $\pm$ 1.4 & & 423.563 & 6.9 $\pm$ 1.4 & & 424.063 & 6.9 $\pm$ 1.4 \\
424.563 & 7.0 $\pm$ 1.4 & & 425.063 & 7.3 $\pm$ 1.5 & & 425.563 & 7.7 $\pm$ 1.5 & & 426.063 & 7.8 $\pm$ 1.6 \\
426.563 & 7.8 $\pm$ 1.6 & & 427.063 & 7.8 $\pm$ 1.6 & & 427.563 & 8.3 $\pm$ 1.7 & & 428.063 & 8.8 $\pm$ 1.8 \\
428.563 & 9.7 $\pm$ 1.9 & & 429.063 & 9.7 $\pm$ 1.9 & & 429.563 & 8.2 $\pm$ 1.6 & & 430.063 & 8.2 $\pm$ 1.6 \\
430.563 & 7.9 $\pm$ 1.6 & & 431.063 & 8.3 $\pm$ 1.7 & & 431.563 & 9.4 $\pm$ 1.9 & & 432.063 & 8.3 $\pm$ 1.7 \\
432.563 & 7.8 $\pm$ 1.6 & & 433.063 & 7.2 $\pm$ 1.4 & & 433.563 & 6.9 $\pm$ 1.4 & & 434.063 & 6.9 $\pm$ 1.4 \\
434.563 & 6.9 $\pm$ 1.4 & & 435.063 & 6.9 $\pm$ 1.4 & & 435.563 & 6.7 $\pm$ 1.3 & & 436.063 & 6.7 $\pm$ 1.3 \\
436.563 & 6.9 $\pm$ 1.4 & & 437.063 & 6.9 $\pm$ 1.4 & & 437.563 & 6.7 $\pm$ 1.3 & & 438.063 & 6.9 $\pm$ 1.4 \\
438.563 & 6.8 $\pm$ 1.4 & & 439.063 & 7.0 $\pm$ 1.4 & & 439.563 & 7.0 $\pm$ 1.4 & & 440.063 & 6.9 $\pm$ 1.4 \\
440.563 & 6.9 $\pm$ 1.4 & & 441.063 & 7.1 $\pm$ 1.4 & & 441.563 & 6.8 $\pm$ 1.4 & & 442.063 & 6.8 $\pm$ 1.4 \\
442.563 & 6.8 $\pm$ 1.4 & & 443.063 & 6.8 $\pm$ 1.4 & & 443.563 & 6.8 $\pm$ 1.4 & & 444.063 & 6.8 $\pm$ 1.4 \\
444.563 & 6.8 $\pm$ 1.4 & & 445.063 & 6.8 $\pm$ 1.4 & & 445.563 & 6.9 $\pm$ 1.4 & & 446.063 & 6.9 $\pm$ 1.4 \\
446.563 & 7.2 $\pm$ 1.4 & & 447.063 & 7.0 $\pm$ 1.4 & & 447.563 & 7.1 $\pm$ 1.4 & & 448.063 & 7.1 $\pm$ 1.4 \\
448.563 & 7.3 $\pm$ 1.5 & & 449.063 & 7.2 $\pm$ 1.4 & & 449.563 & 7.0 $\pm$ 1.4 & & 450.063 & 7.0 $\pm$ 1.4 \\
450.563 & 7.4 $\pm$ 1.5 & & 451.063 & 7.2 $\pm$ 1.4 & & 451.563 & 7.3 $\pm$ 1.5 & & 452.063 & 7.3 $\pm$ 1.5 \\
452.563 & 7.2 $\pm$ 1.4 & & 453.063 & 7.2 $\pm$ 1.4 & & 453.563 & 7.2 $\pm$ 1.4 & & 454.063 & 7.4 $\pm$ 1.5 \\
454.563 & 8.2 $\pm$ 1.6 & & 455.063 & 7.3 $\pm$ 1.5 & & 455.563 & 7.4 $\pm$ 1.5 & & 456.063 & 7.5 $\pm$ 1.5 \\
456.563 & 7.2 $\pm$ 1.4 & & 457.063 & 7.1 $\pm$ 1.4 & & 457.563 & 7.0 $\pm$ 1.4 & & 458.063 & 7.0 $\pm$ 1.4 \\
458.563 & 7.0 $\pm$ 1.4 & & 459.063 & 7.0 $\pm$ 1.4 & & 459.563 & 7.0 $\pm$ 1.4 & & 460.063 & 7.0 $\pm$ 1.4 \\
460.563 & 7.2 $\pm$ 1.4 & & 461.063 & 7.2 $\pm$ 1.4 & & 461.563 & 7.1 $\pm$ 1.4 & & 462.063 & 7.1 $\pm$ 1.4 \\
462.563 & 7.2 $\pm$ 1.4 & & 463.063 & 7.2 $\pm$ 1.4 & & 463.563 & 7.2 $\pm$ 1.4 & & 464.063 & 7.2 $\pm$ 1.4 \\
464.563 & 7.3 $\pm$ 1.5 & & 465.063 & 7.8 $\pm$ 1.6 & & 465.563 & 8.1 $\pm$ 1.6 & & 466.063 & 7.8 $\pm$ 1.6 \\
466.563 & 7.8 $\pm$ 1.6 & & 467.063 & 7.7 $\pm$ 1.5 & & 467.563 & 8.0 $\pm$ 1.6 & & 468.063 & 7.5 $\pm$ 1.5 \\
468.563 & 7.4 $\pm$ 1.5 & & 469.063 & 7.4 $\pm$ 1.5 & & 469.563 & 7.4 $\pm$ 1.5 & & 470.063 & 7.7 $\pm$ 1.5 \\
470.563 & 7.7 $\pm$ 1.5 & & 471.063 & 7.9 $\pm$ 1.6 & & 471.563 & 8.1 $\pm$ 1.6 & & 472.063 & 7.7 $\pm$ 1.5 \\
472.563 & 7.6 $\pm$ 1.5 & & 473.063 & 7.9 $\pm$ 1.6 & & 473.563 & 7.8 $\pm$ 1.6 & & 474.063 & 7.6 $\pm$ 1.5 \\
474.563 & 7.7 $\pm$ 1.5 & & 475.063 & 7.7 $\pm$ 1.5 & & 475.563 & 8.0 $\pm$ 1.6 & & 476.063 & 7.7 $\pm$ 1.5 \\
476.563 & 7.5 $\pm$ 1.5 & & 477.063 & 7.7 $\pm$ 1.5 & & 477.563 & 7.7 $\pm$ 1.5 & & 478.063 & 7.5 $\pm$ 1.5 \\
478.563 & 7.5 $\pm$ 1.5 & & 480.563 & 7.6 $\pm$ 1.5 & & 481.063 & 7.6 $\pm$ 1.5 & & 481.563 & 7.7 $\pm$ 1.5 \\
482.063 & 7.6 $\pm$ 1.5 & & 482.563 & 7.6 $\pm$ 1.5 & & 483.063 & 7.7 $\pm$ 1.5 & & 483.563 & 7.6 $\pm$ 1.5 \\
484.063 & 7.6 $\pm$ 1.5 & & 484.563 & 7.6 $\pm$ 1.5 & & 485.063 & 7.6 $\pm$ 1.5 & & 485.563 & 7.5 $\pm$ 1.5 \\
486.063 & 7.5 $\pm$ 1.5 & & 486.563 & 7.5 $\pm$ 1.5 & & 487.063 & 7.5 $\pm$ 1.5 & & 487.563 & 7.5 $\pm$ 1.5 \\
488.063 & 7.5 $\pm$ 1.5 & & 488.563 & 7.6 $\pm$ 1.5 & & 489.063 & 7.7 $\pm$ 1.5 & & 489.563 & 8.2 $\pm$ 1.6 \\
490.063 & 8.3 $\pm$ 1.7 & & 490.563 & 7.9 $\pm$ 1.6 & & 491.063 & 7.9 $\pm$ 1.6 & & 491.563 & 8.0 $\pm$ 1.6 \\
492.063 & 8.1 $\pm$ 1.6 & & 492.563 & 8.2 $\pm$ 1.6 & & 493.063 & 8.5 $\pm$ 1.7 & & 493.563 & 9.2 $\pm$ 1.8 \\
494.063 & 9.9 $\pm$ 2.0 & & 494.563 & 9.0 $\pm$ 1.8 & & 495.063 & 9.6 $\pm$ 1.9 & & 495.563 & 8.7 $\pm$ 1.7 \\
496.063 & 8.1 $\pm$ 1.6 & & 496.563 & 8.0 $\pm$ 1.6 & & 497.063 & 8.0 $\pm$ 1.6 & & 497.563 & 8.1 $\pm$ 1.6 \\
498.063 & 7.8 $\pm$ 1.6 & & 498.563 & 7.7 $\pm$ 1.5 & & 499.063 & 7.7 $\pm$ 1.5 & & 499.563 & 7.8 $\pm$ 1.6 \\
500.063 & 8.1 $\pm$ 1.6 & & 500.563 & 7.7 $\pm$ 1.5 & & 501.063 & 7.6 $\pm$ 1.5 & & 501.563 & 7.6 $\pm$ 1.5 \\
502.063 & 7.6 $\pm$ 1.5 & & 502.563 & 7.6 $\pm$ 1.5 & & 503.063 & 7.7 $\pm$ 1.5 & & 503.563 & 7.6 $\pm$ 1.5 \\
504.063 & 7.7 $\pm$ 1.5 & & 504.563 & 7.8 $\pm$ 1.6 & & 505.063 & 7.8 $\pm$ 1.6 & & 505.563 & 7.8 $\pm$ 1.6 \\
506.063 & 7.7 $\pm$ 1.5 & & 506.563 & 7.7 $\pm$ 1.5 & & 507.063 & 7.6 $\pm$ 1.5 & & 507.563 & 7.6 $\pm$ 1.5 \\
508.063 & 7.6 $\pm$ 1.5 & & 508.563 & 7.6 $\pm$ 1.5 & & 509.063 & 7.7 $\pm$ 1.5 & & 509.563 & 7.8 $\pm$ 1.6 \\
\cline{1-2}\cline{4-5}\cline{7-8}\cline{10-11}
\caption{First frequency of each half Hz signal frequency band in which we set upper limits and upper limit value for that band.}
\end{longtable}

\clearpage



\end{document}